\documentclass{article}
\usepackage[a4paper, total={6.2in, 9in,margin=1in}]{geometry}

\usepackage[most]{tcolorbox}
\usepackage{lipsum} 
\usepackage{setspace}
\usepackage{scalerel}
\usepackage{hyperref}
\usepackage{fancyhdr}
\usepackage{mdframed} 
\usepackage{multirow}
\usepackage{cuted}
\usepackage{amsmath}
\usepackage{yfonts}
\usepackage{lettrine}
\usepackage{calligra}
\usepackage{afterpage}
\usepackage{amsfonts} 

\usepackage{graphicx}
\usepackage{subcaption}
\usepackage{appendix}
\usepackage{adjustbox}
\usepackage{comment}
\usepackage{tikz}
\usepackage{arydshln}

\newcommand{\skippageno}{18}

\let\oldthepage\thepage
\renewcommand{\thepage}{%
    \ifnum\value{page}=\skippageno\relax
        \phantom{\oldthepage}
    \else
        \oldthepage
    \fi
}

\usetikzlibrary{arrows.meta, positioning, fit, backgrounds, shapes, calc}
\usepackage{titlesec}
\usepackage[
  style=numeric,
  sorting=none, 
  maxnames=5,   
  backend=biber,
  citestyle=numeric-comp
]{biblatex}

\DeclareBibliographyCategory{nonhuman}

\newmdenv[
  backgroundcolor=gray!10, 
  innerleftmargin=10pt,
  innerrightmargin=10pt,
  innertopmargin=10pt,
  innerbottommargin=10pt
  skipabove=\baselineskip,
  skipbelow=\baselineskip
]{darkbox}

\definecolor{customblue}{HTML}{002DFF}

\makeatletter
\providecommand*{\diff}%
    {\@ifnextchar^{\DIfF}{\DIfF^{}}}
        \def\DIfF^#1{%
        \mathop{\mathrm{\mathstrut d}}%
        \nolimits^{#1}\gobblespace}
        \def\gobblespace{%
        \futurelet\diffarg\opspace}
        \def\opspace{%
        \let\DiffSpace\!%
        \ifx\diffarg(%
            \let\DiffSpace\relax
        \else
            \ifx\diffarg[%
                \let\DiffSpace\relax
            \else
             \ifx\diffarg\{%
            \let\DiffSpace\relax
        \fi\fi\fi\DiffSpace}

\title{\vspace{-2cm}To Trade or Not to Trade: \\ An Agentic Approach to Estimating Market Risk Improves Trading Decisions}

\author{
    Dimitrios Emmanoulopoulos$^{1, *}$ \and
    Ollie Olby$^{2}$ \and
    Justin Lyon$^{2}$ \and 
    Namid R. Stillman$^{2, *}$\\[1em]
}

\date{\today}
\fancypagestyle{firstpage} 
{
   \fancyhf{}
   \fancyfoot[L]{\noindent\rule{6.2in}{0.4pt}\vspace{0.2mm}
     \scriptsize{
       \textcopyright \hspace{0.5mm} Barclays 2025 \vspace{0.2mm}\\
       This work is licensed under a \href{https://creativecommons.org/licenses/by/4.0/}
       {Creative Commons Attribution 4.0 International License (CC BY).}  \\ 
       Provided you adhere to the CC BY license, including as to attribution, you are
       free to copy and redistribute this work in any medium or format and remix, transform,
       and build upon the work for any purpose, even commercially.             \\
       BARCLAYS is a registered trade mark of Barclays, all rights are reserved. \\\vspace{1mm}
        $^*$ Corresponding authors: dimitrios.emmanoulopoulos@barclays.com, namid@simudyne.com       
        \\\vspace{1mm}
        $^1$Barclays, $^2$Simudyne
        
     }
   }
}

\begin{document}
\maketitle
\thispagestyle{firstpage}
  \begin{abstract}
    Large language models (LLMs) are increasingly deployed in agentic frameworks, in which prompts trigger complex tool-based analysis in pursuit of a goal. While these frameworks have shown promise across multiple domains including in finance, they typically lack a principled model-building step, relying instead on sentiment- or trend-based analysis. We address this gap by developing an agentic system that uses LLMs to iteratively discover stochastic differential equations for financial time series. These models generate risk metrics which inform daily trading decisions. We evaluate our system in both traditional backtests and using a market simulator, which introduces synthetic but causally plausible price paths and news events. We find that model-informed trading strategies outperform standard LLM-based agents, improving Sharpe ratios across multiple equities. Our results show that combining LLMs with agentic model discovery enhances market risk estimation and enables more profitable trading decisions.
  \end{abstract}


\section{Introduction}
\thispagestyle{firstpage}
\lettrine[lines=2, lhang=0.2, loversize=0.1, findent=1em]{T}{\hspace{0.8em} he} recent development of large language models (LLMs) capable of wider ranging technical tasks has significantly changed the scope of quantitative analysis that can be, at least partially, automated. Applications range from technical analysis of a company's fundamental value, wider market sentiment, factor analysis and most tasks involving some form of natural language processing (NLP) \cite{wang2024applicationnaturallanguageprocessing, nie2024surveylargelanguagemodels}. The implications to trading systems will likely be a dramatic increase in the rate and volume of market insights that can be generated to inform decisions. 

The overall capabilities of LLMs have dramatically increased over the last five years \cite{ferrag2025llm}. This has led to an increase in the number of LLMs available, both as proprietary models from frontier labs or as smaller models with open-weights which can be run locally. Given this, the influence of LLMs on trading decisions is expected to be varied and highly model specific. Early work is starting to compare and benchmark these models in tasks specific to financial applications, such as trading decisions, portfolio optimisation, and market analysis \cite{yu2024finmem, ding2024large, lopez2025can, zhao2024revolutionizing, nie2024survey, li2023large, cao2025deep}. As the number of models increases, and their underlying strengths and weaknesses become more apparent, it is expected that different classes of pre-trained models will be more regularly deployed to achieve certain objectives \cite{plaat2025agentic, ke2025survey}. 

While these objectives are likely to be significantly linked to NLP-based tasks, such as text summarisation, analysis, and generation, recent LLM architectures give early evidence that more complex tasks can also be automated. These LLMs, such as the `o' series from OpenAI or `R1' from DeepSeek, generate `reasoning' tokens which result in the model performing more in-context analysis of the generated output and has lead to improved performance over a number of key evaluation measures \cite{openai2024openaio1card, deepseekai2025deepseekr1incentivizingreasoningcapability}. Though it is widely considered that these LLMs are still significantly weaker than human reasoning at most tasks, they appear to demonstrate sufficient capabilities during in-context chain-of-thought messages that they are collectively referred to as `reasoning' LLMs though the degree to which these LLMs are performing actual reasoning is a topic of active debate\cite{ferrag2025llm, shojaee2025illusion, opus2025illusion}. 

While the application of LLMs to finance tasks is still in its early stages, there is an open question around the extent in which LLMs will impact future trading decisions and whether complex tasks, such as model development, can even be automated. In this research study, we address this question. Specifically, we investigate the performance of LLMs when tasked with the identification of a realistic model of financial time series data and the subsequent impact of their analysis on trading decisions. In this sense, we make a separation of two key components within our agentic framework; the \textit{analyst} and \textit{trader} roles\footnote{Throughout our manuscript, we denote LLM agents such as \textit{builder}, \textit{critic}, \textit{analyast} and \textit{trader}, by italics.}. The \textit{analyst} agents are capable of either calculating risk metrics (referred to as \textit{risk analysts}) or summarising recent news stories for sentiment assessment (\textit{news analysts}). Additionally, and perhaps most importantly, we focus on the development of the \textit{risk analyst} to have sufficient capabilities to attempt model discovery. We structure this process of model discovery as the generation of a stochastic differential equation (SDE) that reproduces the statistics of a historical price path. 

\begin{figure*}[!t]
 \centering
 \begin{minipage}{0.5\textwidth}
   \centering
   \includegraphics[width=\textwidth]{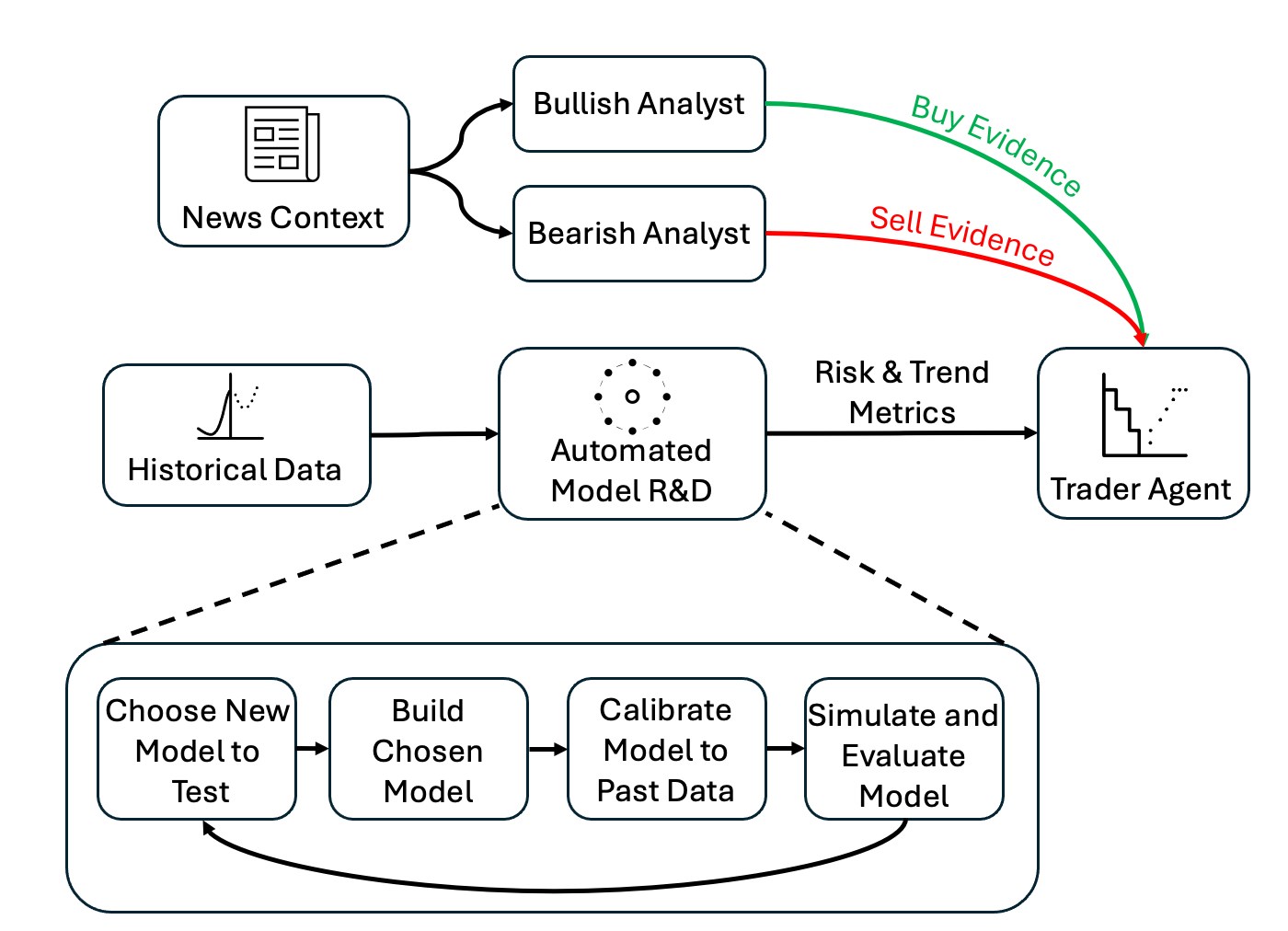}
 \end{minipage}%
 \begin{minipage}{0.5\textwidth}
   \vspace{2em} 
   \caption{\textit{Agentic method for estimating market risk.} We identify a stochastic model of historical price paths using an agentic approach. Here, we follow a \textit{builder}-\textit{critic} pattern for model discovery, in which AI agents implement, simulate, test, and score proposed stochastic models. These models are generated as the output of our agents. We use this to construct and calibrate stochastic models of price paths which provide model-informed estimates for market risk. These risk metrics are provided to a \textit{trader} agent which, along with market context, inform trading decisions. We find that our \textit{risk analyst} agent, using this agentic model discovery loop, improves trading decisions.}
   \label{fig:schematic}
 \end{minipage}
\end{figure*}
Our results show that LLMs are able to perform financial model discovery and refinement, through a process of implementing, calibrating, comparing, summarising, and scoring the proposed stochastic equations. We compare a subset of the latest LLMs on their ability to perform these tasks, highlighting those that are able to identify the stochastic models that best match historical data through a combination of statistical features, symbolic similarity and internal scoring. This agentic framework for model discovery is summarised as the \textit{risk analyst} agent within our \textit{trader}-\textit{analyst} system. 

Given the ability to identify the correct model formulation for the stochastic time series, we next investigate how these agents impact trading decisions, as measured by the impact of their quantification of market risk. To do so, we task a \textit{risk analyst} with discovering suitable stochastic model descriptions for historical equity prices and then estimate market risk statistics (which we refer to as metrics) from these models. These risk metrics are then passed to a trader who can decide to buy or sell depending on these statistics as well as recent information about the stock, as outlined in \autoref{fig:schematic}. We find that our model-informed \textit{risk analyst} improves trading decisions and the degree of improvement is directly connected to the strength of the underlying LLM at model discovery. This highlights that model-informed risk analysis is expected to improve as LLM performance improves. 

Finally, we expect that AI agents will be increasingly dependent on the underlying bias within their training data which consists of historical information up to the so-called knowledge cut-off data (the final date for which the training data relates to). This means that LLMs are likely to be ill-suited for out-of-distribution events that contradict their training data. To test this, we use the \textit{intelligent backtesting} of the \texttt{Simudyne Horizon} market simulator\footnote{SIMUDYNE is a registered trademark of Simudyne Holdings Limited.}. This generates historical time series which matches the statistical properties of the price path but individual per-day shocks relate to synthetic macro-economic events. These results demonstrate that our findings are not dependent on training data, and are robust even if tested on `causally consistent' synthetic data.

Overall, our results demonstrate that currently available LLMs are already sufficiently advanced to augment existing trading decisions with model-informed insights. Furthermore, the process of model discovery, previously assumed to require a high-degree of technical and market-specific knowledge, can be semi-automated in a principled way but we expect that this will continue to require some degree of human supervision. Overall, we find that these insights improve current trading decisions when tested using both traditional and intelligent backtesting methods. Finally, our research introduces new methods for evaluating the performance of LLMs. This form of evaluation highlights a novel new approach to benchmarking LLMs, where underlying LLMs are exposed to real-world tasks as in other evaluation frameworks focused on agentic systems \cite{yao2024tau}. 

\subsection{Related work}

\lettrine[lines=2, lhang=0.2, loversize=0.1, findent=0.5em]{R}{\hspace{0.8em} ecent} work has shown that LLMs display broad capabilities in many natural language tasks. These skills have been leveraged towards more complicated tasks through the use of agentic workflows. 
For example, recent work shows how to employ agentic LLM structures to develop trading strategies \cite{xiao2025tradingagentsmultiagentsllmfinancial}. It uses specialist agents specifically tuned for tasks such as equity research, risk assessment, and strategy formulation to recreate the dynamics of a trading firm. This framework ensures precision and explainability in making financial trading decisions. Alternatively, a novel human-AI interface was developed for alpha mining \cite{wang2023alphagpthumanaiinteractivealpha}. By leveraging LLMs, `Alpha-GPT' interprets quantitative researchers’ ideas, generates formulaic alphas, and iteratively refines them through user feedback. By integrating natural language understanding, reasoning, and structured data synthesis, Alpha-GPT streamlines the discovery, evaluation, and deployment of novel trading signals.

Additionally, other work uses multi-modal LLMs to filter through eleven documents providing background in financial alpha mining strategies \cite{kou2024automatestrategyfindingllm}. This filtered information is then passed to a multi-agent decision making framework to determine seed alphas. The performance from each agent is weighted using deep learning methods to determine the overall strategy. 
This is approach was expanded to consider a novel framework that leverages a mixture of experts (MoE) approach for stock trading using LLMs \cite{ding2024tradexpertrevolutionizingtradingmixture}. It integrates four LLM `experts', each focusing on separate data sources: news articles, market data, alpha factors, and fundamental data. A general expert combines insights from the four experts for final predictions. 

Recently, a layered memory approach was shown to enhance trading performance \cite{li2023tradinggptmultiagentlayeredmemory}. This leveraged LLMs' ability to determine the relevance of a particular piece of information and then store that in long, short or medium term memory. This memory storage framework was proposed to create a more human-like approach to trading, reflecting how individuals retain significant financial events over extended periods. Other work explores a three-step methodology that used LLMs to transform qualitative financial insights into quantitative predictions \cite{lee2024quantifyingqualitativeinsightsleveraging}. The LLM used a moving 5-shot prompting technique to reflect weekly trends and enhance the predictive abilities.

Finally, recent works explored model discovery using LLMs \cite{goodman2024, stillman2024neurosymbolictradersassessingwisdom}. These works used a builder-critic exploration loop, as first set-out in the seminal work by George E. Box \cite{GeorgeBox1962}. This has been extended to financial markets by using vision language models which generate a neural-symbolic model for the fundamental price of an asset \cite{stillman2024neurosymbolictradersassessingwisdom}. Our work builds upon this research by extending their neural-symbolic model discovery framework with LLMs to complement an agentic trading system.

\section{Methodology}

\lettrine[lines=2, lhang=0.2, loversize=0.1, findent=0.5em]{A}{\hspace{0.8em} gentic} AI workflows describe LLMs that use tools to achieve a designated task. In this research, we show how AI agents can be integrated into a model discovery loop to identify stochastic models for simulating financial time series. The resulting insights from the (LLM-discovered) stochastic simulator can inform trading decisions such as by determining model-specific risk metrics. 

All of our experiments were performed on NVIDIA GPUs (V100s, A10s, A100s), for assessing LLM performance at discovering stochastic models. In total, we used $\sim$1,100 GPU hours to calibrate and simulate the proposed stochastic models. All calls to LLMs were made using \texttt{AWS Bedrock}, except for OpenAI which was made using the \texttt{OpenAI API}. For our experiments, we selected a subset of LLMs that were available during the period of January to May 2025. We acknowledge that our selection of language models is neither extensive nor exhaustive, and future work will extend our results to other large and multimodal language models. For data preprocessing, we used NVIDIA \texttt{RAPIDS}\footnote{ \hyperlink{here}{https://github.com/rapidsai/cudf}.}, for simulation and calibration of the stochastic models we used \texttt{diffrax} and \texttt{optax}, respectively \cite{kidger2022neuraldifferentialequations}, on NVIDIA GPUs. For LLM agentic prompt construction, we used \texttt{dspy} \cite{khattab2024dspy}. Finally, for benchmarking LLMs within an advanced backtesting environment, we used \texttt{Simudyne Horizon}, as outlined in \autoref{sec:horizon_results}. 

\subsection{Risk Analyst Agents}

We combine an agentic AI framework with model discovery to identify stochastic models of financial time series. These models are used to simulate financial time series and then estimate model-informed risk statistics, such as value-at-risk (VaR) and maximum drawdown (MDD). These risk metrics are then used to inform an LLM-based trading agent.

\subsubsection{Stochastic Simulator}\label{sec:sde_model}

We are specifically interested in identifying a model, $
m \in \mathcal{M}$, where $m$ is a concrete realisation of the superset of stochastic processes, such as geometric Brownian motion. We constrain the space of all possible models, $\mathcal{M}$, to those that can be written as a stochastic differential equation (SDE), namely
\begin{equation}
\diff S_t(\theta) = f(S_t,t;\theta)\diff t + g(S_t,t;\theta) \circ \diff W_t
\label{eq:SDE_standard}
\end{equation}
in which $f(S_t,t)$ denotes the deterministic term, $g(S_t, t)$ denotes the diffusion term and we assume that $S_t$ is a continuous $\mathbb{R}^n$-valued process representing the evolution of price dynamics of an asset. The stochastic term includes composition with Brownian motion, $\circ$ $\diff W_t$, which we assume to be sampled from the normal distribution with mean zero and standard deviation proportional to the time step, $\diff W_t \sim \mathcal{N}(0, \diff t)$. This model, $m$, is then used to simulate the price dynamics of an asset, $S_t$, given some historical price data, $S^*_t$, and model-specific parameters, $\theta$. 

The functional form of an SDE approximating financial time series is likely to be complex and time varying. This is because exogenous shocks to markets, such as the COVID-19 pandemic, are unlikely to be well-described when using historical data that does not contain these events. This is the `black swan' problem of model selection in finance \cite{ORLIK2024105905}. We note that functions $f$ and $g$ can be multi-dimensional complex functions but, in this work, we restrict the space of SDEs to a single independent variable, considering only the evolution of the price, $S_t$, and not, for example stochastic models of both volatility and price paths. 

Given a proposed model of the market dynamics, $m$, we need to identify the correct parameters, $\theta$ that match model output, $S_t$, to historical data, $S^*_t$. To calibrate and simulate the SDE, we convert the SDE model terms into differentiable compute graphs (using the \texttt{diffrax} package) and calibrate using gradient descent (using the \texttt{optax} package)\cite{kidger2022neuraldifferentialequations}. This reduces the reliance on more complex calibration methods, scales to an arbitrary number of parameters, and allows for a simple loss functions calculated around either summary statistics or mean absolute error (MAE) between outputs. In this work, we calculate the loss, $L$, as the MAE between synthetic and historical price paths, 
\begin{equation}
\mathcal{L}(\theta) = \mathrm{MAE}(S_{t;\theta}, S_{t;\theta}^*) = \frac{\Sigma_i^{N^\mathrm{sims}} |S_{t;i,\theta} - S^*_{t;i,\theta}|}{N^\mathrm{sims}},
\end{equation}
for $N^\mathrm{sims}$ simulation paths and where we look at the time average, comparing per timesteps across Monte Carlo realisations of the SDE. We note that we also considered function-based calibration which minimises, for example, the difference between moments of the time series but this resulted in worse results overall. However, these summary statistics are used when proposing models to test and calibrate during the model discovery loop, as outlined in \autoref{sec:model_disc}. Note, that we have also investigated other statistical properties to evaluate the time series, i.e. the Fourier and signature transform, but found that these dramatically increased the computational time without offering material improvements.

\subsubsection{Model Discovery}\label{sec:model_disc}

\begin{figure*}[!t]
\hspace{1em} \includegraphics[width=\textwidth]{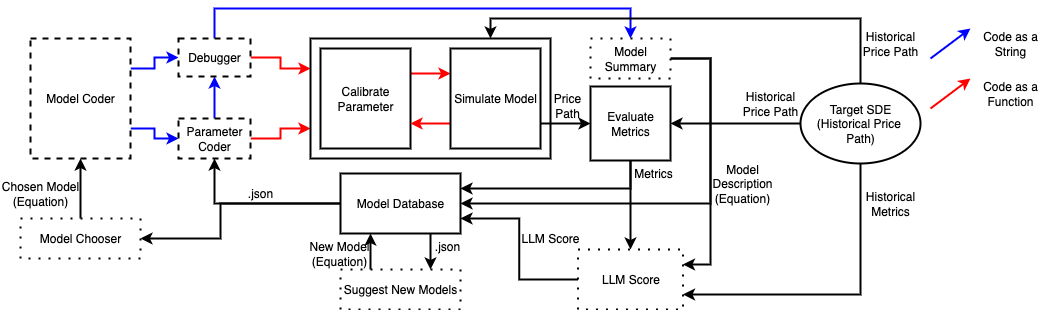}
 \caption{\textit{Risk analyst agentic workflow}. The \textit{risk analyst} uses a \textit{builder} (dashed boxes) and \textit{critic}  (dotted boxes) framework for model discovery, in which the \textit{builder} is tasked with implementation and simulation of the time series. Meanwhile, the \textit{critic} calculates evaluation metrics on the simulated data and uses this to proposed new models to be implemented by the \textit{builder}. }
 \label{fig:analyst_framework}
\end{figure*}

Agentic frameworks refer to the use of systems of LLMs and text-based sub-routines which are designed to complete a pre-determined task (usually complex and multi-step). These sub-routines include function calling, text summarisation, code debugging, and other tasks that LLMs have shown considerable capabilities at in the recent years \cite{basyal2023textsummarizationusinglarge, jiang2025ledextrainingllmsbetter, faldor2025omniepicopenendednessmodelshuman}. Here, we introduce an agentic framework for the discovery of mathematical models of stochastic time series, in which programs for simulating the time series are proposed, implemented, assessed and scored by LLMs. We refer to this as a model-discovery loop in reference to the seminal work by George Box \cite{GeorgeBox1962}, and inspired by \cite{goodman2024}. Our own model-discovery loop is set-out in  \autoref{fig:analyst_framework}.

The process of model discovery follows a typical \textit{builder}-\textit{critic} feedback, where one set of tasks relates to the implementation of the data-generating model, with the aim that the model output is to be representative of historical observations. We refer to this set of tasks as the \textit{builder}. Additionally, the synthetic data are then evaluated and compared to the historical data using a combination of accuracy metrics such as the moments of the time series. Comparisons between the distributions and these are then scored and used to generate model improvements. These tasks are collectively referred to as the \textit{critic}.  

Our aim is to have the \textit{builder} and \textit{critic} collaborate in an `open-ended' environment \cite{hughes2024openendednessessentialartificialsuperhuman}. However, there is a significant technical challenge in giving the LLM unbounded control over both the design and implementation of a stochastic model. To balance structure and flexibility, we constrain the LLM’s invention within an SDE framework, as set out in \autoref{sec:sde_model}. While the framework provides mathematical structure, the LLM itself has full autonomy in suggesting and designing the SDE model. We give more details on both \textit{builder} and \textit{critic} below. 

First, we consider the \textit{builder}. Here, the primary goal is to convert a specific SDE equation into a working calibrated model that can generate appropriate historical data. The \textit{builder} can be expressed as
\begin{equation}
\mathcal{B}:(m_n, c_n) \xrightarrow{}F_n(t;\theta) = S_{t,n},
\label{eq:builder}   
\end{equation}
where the candidate model, $m_n \in \mathcal{M}$, is taken from the space of proposed models, $\mathcal{M}$, and where $c_n \in \mathcal{C}$ is the context for choosing this model, for $n = 1, ..., N$ iterations of the model-discovery loop. We set the context as the history of previously proposed and tested models which we refer to as the \textit{memory}. Additionally, $F_n(t; \theta)$ is the program implementing the model, $m_n$, given appropriate parameters, $\theta$ and which generates the stochastic price paths, $S_{t,n}$. The \textit{builder} includes the following sub-tasks:
\begin{enumerate}
\item{Implementing: Given the proposed model, $m_n$, the `implementer' constructs a simulator, $F_n(t;\theta)$, which generates synthetic stochastic time series. We use a reference implementation as context for the implementer (using the \texttt{diffrax} package). We note that this reference implementation reproduces geometric Brownian motion, which is the first starting model used in the model-discovery loop, and the reference is static through subsequent iterations. The implementer includes both the initial construction of the program, and a parameter estimation step in which the model suggests an initial value, $\theta^0$, for the model calibration.}
\item{Simulating: Given the simulator, $F_n(t; \theta)$ and initial estimates for parameters, $\theta^0$, we then calibrate the model using the historical data, as outlined in \autoref{sec:sde_model}, to determine parameters, $\theta$, and use this to generate Monte Carlo realisations of the stochastic time series, $S_{t,n}$, approximating the historical price paths.}
\end{enumerate}

Following the construction of the model, the synthetic data are then passed to a second sub-routine which we refer to as the \textit{critic}, expressed by

\begin{equation}
\mathcal{R}:(S_{t,n}, c_n)\xrightarrow{}c_{n+1}, m_{n+1},
\label{eq:critic}
\end{equation}
where $S_{t,n}$ is the set of Monte Carlo realisations of the price paths generated from the simulator; $c_n$ is the current agent context, which involves all previously tested models and their evaluation metrics, $c_{n+1}$ and $m_{n+1}$ are the updated context and the chosen next model to test. The \textit{critic} includes the following tasks: 
\begin{enumerate}
\item{Testing: We calculate the evaluation metrics of the synthetic data, $e(S_t, S^*_t)$. These include summary statistics of the historical and simulated time series such as statistical moments (mean, variance, kurtosis, skew), tail metrics for the 95 percentile, the Hill estimator, Hurst exponent, jump intensity, autocorrelation function (ACF) metrics, growth rate, and the mean absolute deviation from the median \cite{tsay2005analysis, cont2003financial}. These statistics were chosen because they carry a large degree of information about the features of the generated time series and are known to be of relevance for financial time series\cite{safari2025international, ratliff2023revisiting}. Alongside this, we consider direct tests between the distribution of returns of historical and synthetic time series for which we calculate the Kolmogorov-Smirnov test, Wasserstein distance, and the mean absolute percentage error \cite{hyndman2018forecasting, villani2008optimal, hollander2013nonparametric}. Finally, we use an additional agent to construct a summary of the tested model, $m_n$, used to generate time series, $S_{t,n}$.} 
\item{Scoring: The above evaluation metrics are then passed to a scoring agent, which provides an overall score based on both the loss of the metrics, $\mathcal{L}(\theta)$, and a novelty score, $K_n$. The latter, is derived from an additional agent, using the model summary to determine a score between 1-100 on the novelty or `interesting-ness' of the mathematical equation. This follows previous work that highlights that LLMs are potentially approaching human-like effectiveness at determining `interesting-ness' due to the human biases implicit in their training data \cite{Chiang2023CanLL, faldor2025omniepicopenendednessmodelshuman, hughes2024openendednessessentialartificialsuperhuman}. While we note that a simple scoring function can be estimated, which weights different components of the evaluation metrics to balance, for example, novelty versus fidelity, we consider a purely agentic method here to remove any reliance on tuning hyper-parameters and make a more principled comparison against LLMs. The model context is then updated to the model context, $c_{n+1} \in \mathcal{C}$, to determine the next model for testing.}

\item{Suggesting: Given the list of proposed models $\mathcal{M}$, their calibration loss, $\mathcal{L}(\theta)$, evaluation metrics, $e$, and novelty score, $K$, an LLM is then tasked with generating new candidate models that are expected to improve on previous iterations. This is based on the assumption that LLMs are able to approximate human creativity through their pre-training \cite{kumar2024human, tian2023macgyver}. These candidate models are then appended to the candidate model space.}

\item{Choosing: From the space of candidate models $\mathcal{M}$, the critic determines the next model to pass to the builder for implementation, $m_{n+1}$. This choice is based on the context $c_n$ of previously tested models.}

\end{enumerate}

The output of the \textit{critic} loop is the next candidate model for the \textit{builder} to implement and simulate. The model discovery process then repeats for $N$ iterations. Our model discovery loop involves repeated calls to LLMs which are tasked with specific agentic sub-routines. The orchestration of agents as either \textit{builders} or \textit{critics} is achieved using \texttt{DSPy} as the general framework for implementing agentic calls \cite{khattab2024dspy, khattab2022demonstrate}. Example outputs from the model-discovery loop agents are given in \autoref{app:prompt}.

\begin{figure*}[t!]
\centering
\begin{subfigure}{0.29\textwidth}
 \centering
 \includegraphics[width=\linewidth]{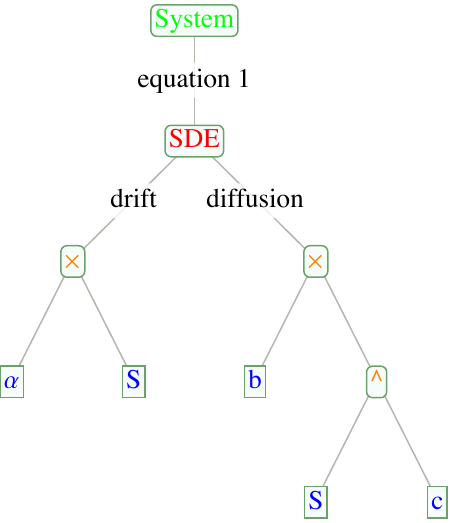} 
 \caption{\footnotesize{$\diff S_t = aS_t \diff t + bS_t^c \diff W_t$}}
 \label{fig:CEV_tree}
\end{subfigure}
\hfill
\begin{subfigure}{0.29\textwidth}
 \centering
 \includegraphics[width=\linewidth]{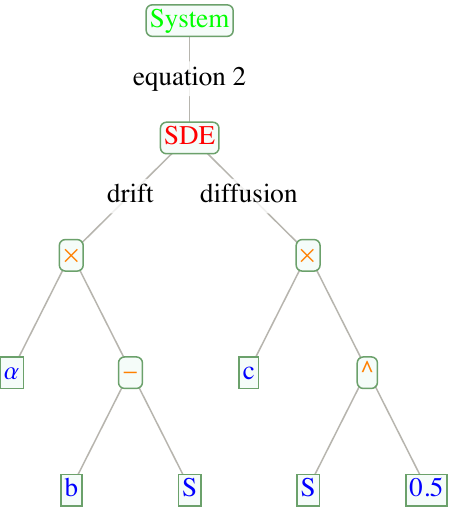}
 \caption{\footnotesize{$\diff S_t=a(b - S_t)\diff t + c\sqrt{S_t} \diff W_t$}}
 \label{fig:CIR_tree}
\end{subfigure}
\hfill
\begin{subfigure}{0.29\textwidth}
 \centering
 \includegraphics[width=\linewidth]{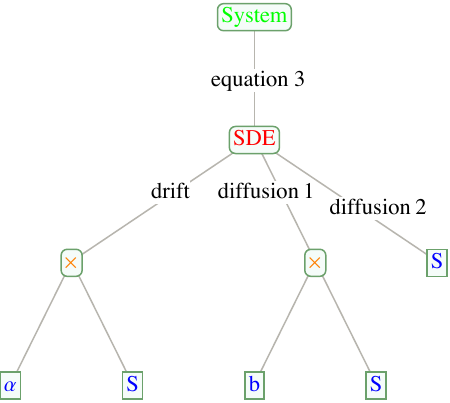}
 \caption{\footnotesize{$\diff S_t = aS_t \diff t + bS_t \diff W_t + S_t dJ_t$}}
 \label{fig:jump_tree}
\end{subfigure}
\caption{\textit{Symbolic trees.} We convert the stochastic differential equation, used to simulate the financial time series, into a directed acyclic symbolic tree which we then use to calculate symbolic similarity and diversity scores. Here, we show (a) the CIR model, (b) the CEV model, and (c) the JD model, which are described in \autoref{sec:model_disc_results}.}
\label{fig:sym_score}
\end{figure*}

\subsubsection{Symbolic Similarity Score}\label{sec:sym_trees}

In order to quantify the degree to which our model discovery loop is able to correctly identify a mathematical model, we construct a symbolic similarity score, $K_\mathrm{similarity}$. Thus, we convert the proposed symbolic expression, describing model $m \in \mathcal{M}$, in to a directed acyclic graph in which each node represents the variable or parameter and edges represent the mathematical operation. We further decompose the model into stochastic and deterministic components, representing the drift and diffusion terms of the underlying SDE. \autoref{fig:sym_score} shows the symbolic trees for the three target equations discussed in \autoref{sec:model_disc_results}.

Given a symbolic tree, we calculate graph-based statistics such as the mathematical complexity (the average number of paths within the tree) and the Weisfeiler-Lehman (WL) graph hash (a simple graph-based representation of the equation \cite{shervashidze2011weisfeiler}). We then use the WL graph hash to calculate the similarity between two graphs, $K_\mathrm{similarity}$, given by the following equation:

\begin{equation}
    K_\mathrm{similarity} = K_\mathrm{WL}(G_i, G_j),
\end{equation}
in which $G$ is the symbolic graph generated from the proposed and tested model, $m$, and $K_\mathrm{WL}$ is the WL graph kernel calculated from WL graph hashes. The diversity of the proposed stochastic models, discovered by our \textit{risk analyst} agents, $K_\mathrm{diversity}$, is calculated as one minus the average pairwise similarity of all unique symbolic graph pairs, generated from the tested models, $m_n$, namely

\begin{equation}\label{eq:diversity}
    K_\mathrm{diversity} = 1 - \frac{1}{\binom{N}{2}} \sum_{1 \leq i < j \leq N} K_\mathrm{WL}(G_i, G_j),
\end{equation}
for $N$ models generated by the model discovery loop. Finally, we also calculate the overall complexity of the symbolic equation, a proxy for model complexity, $K_\mathrm{complexity}$. This is defined as the average path length for the tree, in which larger equations with more terms and mathematical operations result in a higher complexity score. We next outline how the risk metrics are calculated from the final chosen model. These metrics are used by the \textit{trader} agents to determine trading decisions, discussed in \autoref{sec:traders}.

\subsubsection{Model-informed Risk Metrics}

The above model-discovery loop is used to determine a mathematical model for simulating the stochastic process that best matches a historical financial time series. Stochastic processes are widely used in financial trading applications, including option pricing, portfolio optimisation, and quantifying market risk \cite{rolski2009stochastic, kijima2002stochastic}. In this research study, we focus on market risk analysis, using the final model proposed by the model-discovery loop (\autoref{sec:model_disc}) to calculate particular market risk statistics. Specifically, we calculate the value-at-risk (VaR), conditional value-at-risk (CVaR) and maximum drawdown (MDD), which are fundamental to risk management in financial markets \cite{Zhang_2013}, \cite{Rockafellar2000OptimizationOC}.

Given the generated model of the price path of an asset, denoted by $S_t$, we consider the loss to be $L_t = S_0 - S_t$, VaR is then a measure of these potential losses, defined as the quantile of the loss distribution for a given confidence level $(1-\alpha)$, 
\begin{equation}
\mathrm{VaR}_\alpha = \inf\{l : P(L_t\leq l)\geq \alpha\}.
\label{eq:var_loss_meas}
\end{equation}
The CVaR, calculates the expected loss, given that we exceed VaR,
\begin{equation}
\mathrm{CVaR}_\alpha = E[L_t|L_t>\mathrm{VaR}_
\alpha]
\label{eq:cvar_exp_sho_fal}
\end{equation}
giving the magnitude of expected losses, rather than only the threshold.

Alongside VaR and CVaR, the MDD of a strategy over some period $[0,T]$ is commonly used to identify the relative maximum losses (peak-to-trough) incurred,

\begin{equation}
\mathrm{MDD} = \max_{0\leq t \leq T}\left[\frac{\max_{0\leq r \leq t}\left(S(r) - S(t)\right)}{\max_{0\leq r \leq t}S(r)}\right].
\label{eq:mdd}
\end{equation}

These measures are fundamental in estimating risk exposure, but are known to underestimate inherent tail risk \cite{embrechts2011quantitative}. Given an appropriate model of market prices, Monte Carlo (MC) sampling can be used to  to generate $N^\mathrm{sims}$ price paths from the SDE. These paths will be used to calculate both VaR and CVaR to approximate the tail risk.

Finally, we also use extreme value theory (EVT) to directly estimate the model- and market- specific tail risk. We consider the residuals, $\epsilon_t$, and the price paths to quantify model adequacy and to imply tail behaviour, respectively. For the residuals, given an appropriately calibrated SDE, we extract the standardized residuals as
\begin{equation}
\epsilon_t = \frac{S_{t + \Delta t} - S_t - f(S_t, t)}{g(S_t,t)\sqrt{\Delta t
}}
\label{eq:stand_resid}
\end{equation}
for drift and diffusion terms, $f(S_t,t)$ and $g(S_t, t)$, respectively. 

For a value of interest, $X$, we calculate the exceedances given a pre-defined threshold, $u$. Hence, we follow the peaks-over-threshold method and fit a generalized Pareto distribution (GPD) to the exceedances \cite{coles2001introduction, mcneil2015quantitative},
\begin{equation}
P(X>x | X>u) = \left(1 + \xi \frac{x-u}{\gamma} \right)^{-1/\xi},
\label{eq:gpd}
\end{equation}
in which $\xi$ is the shape parameter, for which we assume $\xi \neq 0$, and $\gamma$ is the scale parameter. The shape parameter gives estimate for the $k^\mathrm{th}$ moment, given $\xi < 1/k$, and the scale parameter denotes the dispersion of these exceedences. By applying EVT to both residuals and the price paths, these parameters provide insights into the quality of the model fit to the historical market dynamics and the model-implied extreme behaviours.

When applying EVT to the simulation paths directly, we calculate the losses, $L_t$, and then fit the conditional excess above these losses using a threshold of 95\% of the empirical loss distribution. For residuals, this calculates the extent to which our simulator has correctly reproduced the tail risk for the historical market price paths. Alternatively, for the losses of the simulated price paths, we obtain a better estimate for the tail risk than would be possible using historical data only, by sampling over Monte Carlo realisations. 

We note that the above implementation assumes that the excess losses are independent which is typically not the case for financial time series. Further steps, such as through de-clustering, could be introduced to improve upon the risk analysis described above. However, we restrict our results to the above simplified implementation due to the challenges in automating more advanced risk analysis which we set as a topic for future research. We next set out how these measures are incorporated into trading strategies.

\subsection{Trader Agents}\label{sec:traders}

Having determined a stochastic model of the time series, we then use the stochastic simulator to generate an array of price paths which are used to quantify market risk, and inform a trading strategy. In order to construct a trading strategy, we take an agentic approach and construct \textit{trader} agents which make trading decisions based on model-informed risk and trend metrics, as well as market context taken from recent news headlines. The \textit{trader} agents consist of a separate series of sub-routines, executed by an LLM. Here, we focus on equity markets, considering individual stocks. However, we note that our approach is broadly applicable to any asset class. We outline our methodology for combining market risk and price trends in a trading strategy below.

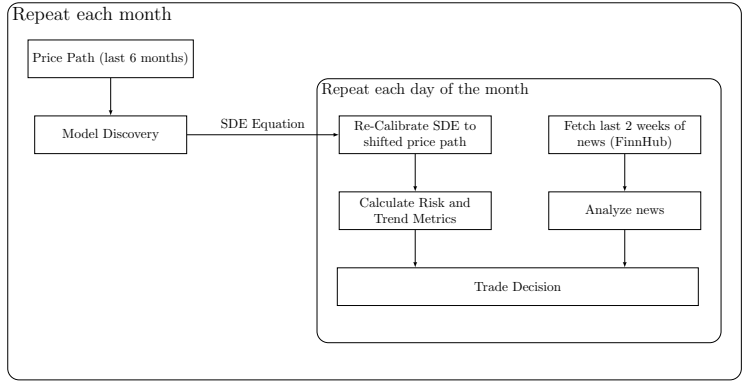
\begin{figure}
    \centering
    \scalebox{0.5}{
    \begin{tikzpicture}[
        node distance = 1cm and 1.5cm,
        box/.style = {rectangle, draw, minimum width=4cm, minimum height=1cm, align=center},
        boxRound/.style = {rectangle, draw, minimum width=4cm, minimum height=1cm, align=center, rounded corners=3mm},
        wide box/.style = {rectangle, draw, minimum width=9.5cm, minimum height=1cm, align=center,node distance=10pt},
        container/.style = {rectangle, draw, rounded corners=3mm, align=center, minimum height=7cm},
        maincontainer/.style = {rectangle, draw, rounded corners=3mm, align=center, minimum height=10cm},
        >={latex}
        ]  
        \node[box] (price) {Price Path (last 6 months)};
        \node[box, below=of price] (model) {Model Discovery};
        \node[box, right=4cm of model] (recalibrate) {Re-Calibrate SDE to\\shifted price path};
        \node[box, right=of recalibrate] (news) {Fetch last 2 weeks of\\news (FinnHub)};
        \node[box, below=of recalibrate] (risk) {Calculate Risk and\\Trend Metrics};
        \node[box, below=of news] (analyze) {Analyze news};
        \node[wide box, below=1 cm of risk, xshift=2.7cm] (trade) {Trade Decision};   
        \draw[->] (price) -- (model);
        \draw[->] (model) -- node[above] {SDE Equation} (recalibrate);
        \draw[->] (recalibrate) -- (risk);
        \draw[->] (news) -- (analyze);
       
        \draw[->] (risk) -- (risk |- trade.north);
        \draw[->] (analyze) -- (analyze |- trade.north);
       
        \begin{scope}[on background layer]
            \node[container, fit=(recalibrate)(news)(risk)(analyze)(trade), inner sep=15pt] (daily) {};
            \node[anchor=north west] at (daily.north west) {\large{Repeat each day of the month}};
           
            \node[maincontainer, fit=(price)(model)(daily), inner sep=15pt] (monthly) {};
            \node[anchor=north west] at (monthly.north west) {\Large{Repeat each} month};
        \end{scope}
    \end{tikzpicture}
    }
    \label{fig:pipeline}
\caption{\textit{Overview of the trading strategy pipeline.} Each month, the model discovery phase identifies a suitable SDE based on the past six months of price data. On each trading day, the SDE is recalibrated to the latest market conditions, recent news is fetched via the FinnHub API. Risk, trend metrics and sentiment are analyzed. These inputs are then used to inform a daily trade decision.}
\label{fig:traderframework}
\end{figure}

\subsubsection{Trend Metrics}
\label{sec:trend_metrics}

In order to give the LLM enhanced context for making trades, we use the relative strength index (RSI) to show the current trend of the price path \cite{chio2022comparativestudymacdbasetrading}. RSI is a measure of signal momentum and is commonly used for enhancing trading decisions by validating trends and trend reversals. It is defined as
\begin{equation}
\mathrm{RSI} = 100-\left(\frac{100}{1+\frac{\text{Average Gain}}{\text{Average Loss}}}\right).
\label{eq:rsi}
\end{equation}
An RSI value below thirty usually indicates an oversold stock whereas a value above seventy indicates an overbought stock \cite{chio2022comparativestudymacdbasetrading}. We set the lookback period for the loss and gain to fourteen days, as outlined in  \cite{chio2022comparativestudymacdbasetrading}.

Given that the model discovered contains a drift term, it would be useful to exploit this to determine the trend of the signal. The drift term is separated from the SDE and the current price is passed to it. The standard SDE form is expressed in \autoref{eq:SDE_standard}. The drift term represents the expected rate of change of price, $S_t$. If the value of the drift is positive/negative then the price is expected to increase/decrease. This is passed to the LLM as a string for either `positive' or `negative'. We call this the $\mathrm{\text{Drift Polarity}_t}$ term,

\begin{equation}
\text{Drift Polarity}_t = \begin{cases}
\text{`Positive'}, & \text{if } f(S_t,t) \ge 0 \\
\text{`Negative'}, & \text{otherwise }
\end{cases}
\label{eq:drif_pol}
\end{equation}

\subsubsection{Trading Strategies}
\label{sec:trading_strategies}

We have set out how LLMs can be used in an agentic framework for model discovery with a focus on the determination of a stochastic process. The generation of the SDE attempts to reproduce historical price dynamics. The overall class of SDEs is unconstrained and the individual model is determined by a \textit{builder}-\textit{critic} loop. We next set out how the outcome of this loop, the proposed SDE model, is used in a trading strategy. 

To determine the role and impact of LLMs in trading applications, we consider a purely generative approach to trading decisions. To do so, we pass the calculated risk and trend metrics (VaR, CVaR, MDD, EVT metrics, RSI and Drift Polarity) to an LLM that is instructed to make trading decisions. We call this the \textit{trader} agent and the framework for this is provided in Figure~\ref{fig:traderframework}. LLMs are well-known for their ability to take insights from large context windows \cite{ding2024longropeextendingllmcontext}. To take advantage of this we also provide context in the form of news articles to the trading agent and the methodology for this is outlined in \autoref{sec:context}.

At the beginning of each month, we run the model discovery loop and the best SDE (determined by lowest loss) is chosen as the optimal model used for the next twenty days of trading. Here, we restrict the number of times we perform model discovery to monthly due to computational constraints. However, we found that the suggested models were relatively static over multiple consecutive trading days. Future work will look to quantify the consistency of the proposed models over time. 

For each day of the month, we then recalibrate the SDE model to the price path of the last hundred trading days. The risk and trend metrics are then calculated from this calibrated SDE, as set out in \autoref{sec:trend_metrics}. The summary of the news context is constructed by an LLM, which we refer to as the \textit{news analyst}, described in \autoref{sec:context}. This summary is then passed to the \textit{trader}, who makes the final decision to buy, sell, or hold. Example output from the \textit{trader} is given in \autoref{app:prompt}.

The \textit{trader} generates a trade signal, $T_t$, which can take one of three values: buy, sell, or hold. Based on this signal, the portfolio position, $P_t$, at time $t$ is updated according to the following rules:

\begin{equation}
P_t =
\begin{cases}
    \frac{C_t}{S_t}, & \text{if } T_t = \text{buy} \text{ and } C_t > 0 \\
    0, & \text{if } T_t = \text{sell} \\
    P_{t-1}, & \text{if } T_t = \text{hold}
\end{cases}
\label{eq:port_pos}
\end{equation}
for $C_t$, the amount of cash at time $t$. $C_t$ is updated based on execution and transaction costs, such that

\begin{equation}
C_t =
\begin{cases}
    C_{t-1} - P_t S_t - \text{TC}_t, & \text{if } T_t = \text{buy} \\
    C_{t-1} + P_{t-1} S_t - \text{TC}_t, & \text{if } T_t = \text{sell} \\
    C_{t-1}, & \text{if } T_t = \text{hold}
\end{cases}
\label{cash_amnt}
\end{equation}
in which the transaction cost, $\mathrm{TC}_t$, is determined by

\begin{equation}
\mathrm{TC}_t = \kappa  P_tS_t
\label{eq:trans_cst}
\end{equation}
for an assumed transaction charge, $\kappa$, given in basis points (bps). NYSE stocks are estimated to have an average  charge of 8.8 bps of the total value of the trade, such that $\kappa = 8.8 \times 10^{-4}$\cite{Engle2006MeasuringAM}.

Finally, we compare the LLM trading performance to a buy-and-hold strategy. While the buy-and-hold strategy is remarkably simple, it remains highly robust for capitalising on a bullish market, as we report in \autoref{sec:trading_results}.

\subsubsection{Trading Context}
\label{sec:context}
Over medium to long-term time horizons, a variety of other factors inform trading decisions ranging from macro-economic events, the performance of associated assets and companies, news sentiment, etc. To reproduce the effect of these factors on trading decisions, we pass market context to our \textit{trader} agent. 

Here, we use the \texttt{FinnHub} API\footnote{Further information can be found \hyperlink{https://finnhub.io/docs/api}{here}.} to gather context by retrieving news for the selected company within the specified date range. As outlined in \autoref{sec:traders}, we consider trading over several months with trading decisions made daily. Given this, we use news context from the last five days. The news context provided to the \textit{trader} is designed to give the agent access to the breadth of detail around current events. The \texttt{FinnHub} API summarises the news article into a short paragraph. This summary is key to keeping the context window short while still conveying the core message of the article. Each news item includes the timestamp of publication, the source of the article, the headline, a brief summary and the company that this item applies to. 

We note that the movement of a company's stock price is typically highly dependent on the progress of other related companies. To account for this, we construct a \textit{news analyst} agent that collects news around a specific company of interest by identifying other companies that may affect the price of the company in question. We then retrieve news for these related companies, with the assumption that this additional information of the wider sector/industry will improve trading decisions.

\begin{figure}[!t]
 \centering
 \includegraphics[width=0.9\textwidth]{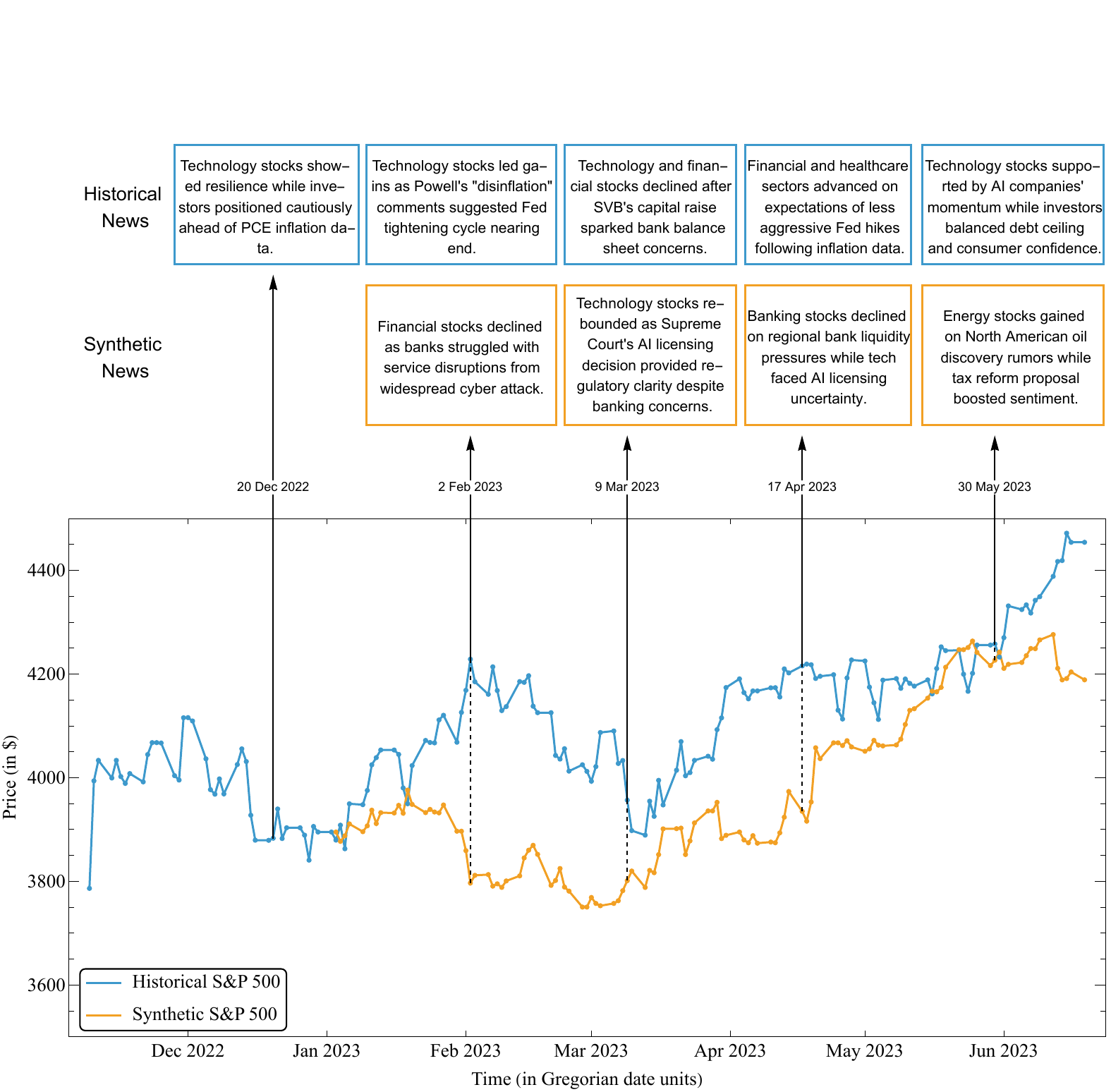}
 \caption{\textit{Synthetic news generation}. To ensure that the tested LLMs do not have relevant context in their pretrained weights to improve trading results, we use the \texttt{Simudyne Horizon} simulator to generate synthetic price and news data. This provides a more advanced form of context-aware backtesting where trading strategies are tested on different plausible news cycles. In the \autoref{app:horizon_context} we give the full output of the generated news articles.}
 \label{fig:synthetic_news}
\end{figure}

The \textit{news analyst} agent determines the pros and cons to investing in this particular asset, according to the list of news items, and provides a summary of context-specific insights. Finally, these insights are passed to the \textit{trader} to inform their trading decision. Example output from the \textit{news analyst} is given in \autoref{app:prompt}.

\subsubsection{Synthetic Trading Context}

In the above description, we have set out a strategy of implementing and improving stochastic models for quantifying market risk using LLMs. We use these stochastic models to calculate risk and trend metrics that are passed to \textit{trader} agents to determine trading (buy/sell/hold) decisions. However, when testing the performance of the \textit{trader} agents, we could not preclude the possibility that historical events had, under some circumstances, been accessible to the pre-trained LLMs, despite all testing periods being outside of the models knowledge cut-off dates.

To rule out this biasing we use \texttt{Simudyne Horizon}\footnote{~\href{https://www.simudyne.com/solutions/simudyne-horizon/}{\texttt{Simudyne Horizon}} is a hybrid market simulator combining traditional interpretable methods for simulating financial time series with agentic workflows.~\texttt{Simudyne Horizon} generates synthetic time series and news data across capital markets, reproducing realistic price dynamics over days to years, while also providing interpretable estimates for the trading activity as grouped by trading styles. \\} to generate synthetic time series which have the statistical properties of historical price paths but relate to a set of synthetic market events. These events are described through news stories generated by LLMs, which blend real historical events with fictional ones by aligning recent price changes to plausible but fictional news stories such as cyber-attacks, alternative policy announcements, or other geopolitical/market events. This means that trading decisions informed by both price data and market context (such as when driven by sentiment analysis) can be effectively tested over a range of extreme scenarios without putting capital at risk. We show an example of the \texttt{Simudyne Horizon} context-aware backtesting functionality in \autoref{fig:synthetic_news} and give examples of fictional headline summaries in \autoref{app:horizon_context}. 

We use this as a testing environment to evaluate our \textit{trader} agents, ensuring they cannot be influenced by data from their pre-training even when events are outside of their knowledge cut-off date. This bias may occur, for example, when events have a high likelihood of occurring due to past historical events, such as conservative estimates around rate decisions (i.e., the decision by central banks to typically make small changes to interest rates). 

\section{Results \& Discussion}
\lettrine[lines=2, lhang=0.1, loversize=0.1, findent=0.5em]{D}{\hspace{0.4em} espite} widespread attention to LLMs, there has been relatively little research on whether they can be used extensively for discovering financial models. Below, we outline our results in applying agents to both the discovery of stochastic models and the application of them to inform trading decisions. Firstly, we outline results comparing seven of the frontier LLMs that represent state-of-the-art performance at the time of writing for their ability to perform automated model discovery. Secondly, we evaluate the LLMs performance as risk-informed agentic traders, considering periods which we expect to be out of the LLMs training data. Lastly, we test the LLMs performance using a context-aware backtesting environment generated by \texttt{Simudyne Horizon}. 

\begin{table}[!t]
\centering
\small
\begin{adjustbox}{width=\textwidth,totalheight=\textheight,keepaspectratio}
\begin{tabular}{|l|cccc|cc|}
\hline
LLM & Loss ($\mathcal{L}(\theta)$) & $K_\mathrm{Score}$ & $K_\mathrm{similarity}(m_i, m^*)$ & Combined Score & $K_\mathrm{diversity}$ & $K_\mathrm{complexity}$\\ 
\hline
\multicolumn{7}{|c|}{Equation 1: $\diff S_t = aS_t \diff t + bS_t^c \diff W_t$} \\
\hline
Deepseek: R1 & \underline{1.0 (±1.5)} & \underline{52 (±8.4)} & \underline{0.44 (±0.03)} & \underline{1060 (±110)} &\underline{0.72} &3.9 \\
Anthropic: Sonnet-3.7 & \textbf{0.040 (±0.043)} & \textbf{65 (±1.8)} & 0.40 (±0.069) & \textbf{1240 (±220)}  &0.67 &4.3 \\
Anthropic: Sonnet-3.5v2 & 60 (±34) & 50 (±3.9) & \underline{0.44 (±0.076)} & 980 (±300) &0.7 &4.0 \\
OpenAI: 4o-mini & 75 (±0.10) & 32 (±2.1) & 0.25 (±0.030) & 390 (±20)  &0.69 &\underline{4.4}\\
OpenAI: o1-mini & 75 (±1.0) & \underline{52 (±4.5)} & 0.33 (±0.041) & 680 (±130) &\textbf{0.88} &3.6\\
OpenAI: o3-mini & 30 (±41) & \underline{52 (±5.7)} & 0.40 (±0.10) & 880 (±180) &0.68 &3.4\\
Meta: llama-3.3 & 60 (±34) & 36 (±4.6) & \textbf{0.45 (±0.017)} & 690 (±54) &0.66 &\textbf{4.7}\\
\hline
\multicolumn{7}{|c|}{Equation 2: $\diff S_t = a(b - S_t)\diff t + c\sqrt{S_t} \diff W_t$} \\
\hline
Deepseek: R1 & 0.33 (±0.19) & \underline{64 (±2.2)} & 0.35 (±0.060) & \underline{980 (±100)} & 0.72 & 3.8\\
Anthropic: Sonnet-3.7 & \textbf{0.00058 (±0.00048)} & \textbf{76 (±3.3)} & 0.37 (±0.042) & \textbf{1250 (±230)} & 0.67 & \bf{4.4}\\
Anthropic: Sonnet-3.5v2 & \underline{0.26 (±0.23)} & 56 (±3.1) & \underline{0.41 (±0.054)} & 970 (±30) & 0.67 & 3.9\\
OpenAI: 4o-mini & 0.36 (±0.16) & 51 (±2.2) & 0.37 (±0.042) & 820 (±82) & \underline{0.77} & 3.8\\
OpenAI: o1-mini & 0.39 (±0.12) & 58 (±8.0) & 0.34 (±0.060) & 770 (±300) & \textbf{0.86} & 3.8\\
OpenAI: o3-mini & 0.42 (±0.039) & 61 (±8.2) & \textbf{0.42 (±0.093)} & 970 (±170) & 0.59 & 3.2 \\
Meta: llama-3.3 & 0.40 (±0.097) & 54 (±2.2) & 0.27 (±0.097) & 600 (±110) & 0.48 & \underline{4.1}\\
\hline
\multicolumn{7}{|c|}{Equation 3: $\diff S_t = aS_t \diff t + bS_t \diff W_t + S_t \diff J_t$} \\
\hline
Deepseek: R1 & 0.21 (±0.01) & \underline{64 (±1.3)} & 0.42 (±0.17) & 1180 (±280) & 0.7 & \underline{3.9} \\
Anthropic: Sonnet-3.7 & \textbf{0.0064 (±0.010) }&\textbf{69 (±2.5)} & 0.42 (±0.18) & \textbf{1270 (±470)} & \underline{0.72} & \textbf{4.2}\\
Anthropic: Sonnet-3.5v2 & \underline{0.078 (±0.072)} & 63 (±1.3) & \underline{0.47 (±0.14)} & \underline{1190 (±170)} & 0.66 & \underline{3.9} \\
OpenAI: 4o-mini & 0.21 (±0.011) & 46 (±2.2) & \textbf{0.48 (±0.16)} & 720 (±87) & 0.7 & \underline{3.9} \\
OpenAI: o1-mini & 0.21 (±0.00010) & \underline{64 (±4.4)} & 0.34 (±0.21) & 870 (±520) & \textbf{0.83} & \underline{3.9} \\
OpenAI: o3-mini & 0.22 (±0.0048) & \textbf{69 (±4.2)} & 0.45 (±0.15) & \textbf{1270 (±210)} & 0.67 & 3.7\\
Meta: llama-3.3 & 0.21 (±0.0016) & 49 (±5.3) & 0.33 (±0.23) & 800 (±500) & 0.38 & 3.4\\
\hline
\end{tabular}
\end{adjustbox}
\caption{Performance metrics for three different equations, in which $m^*$ relates to each of the ground-truth equations (1-3) for each subsection of the table (\textbf{Bold}=best, \underline{underlined}=second best). Note that for all the metrics apart from Loss, larger values equate to a higher performance.}
\label{tab:equations_performance}
\end{table}

\subsection{Discovering Stochastic Models}\label{sec:model_disc_results}
In order to quantify the degree to which our model discovery loop is effective at identifying reasonable candidate functions for risk analysis, we first test our framework on known SDEs. We consider three SDEs that are commonly used in finance and economics. These are the constant elasticity of variance (CEV) model, the Cox-Ingersoll-Ross (CIR) model, and a simple jump-diffusion (JD) model \cite{beckers1980constant,overbeck1997estimation,cheang2012modern}. We give more details of each below. For each model, we use the same lowercase Roman alphabet for parameter variables, i.e. $a, b, c, \ldots$, for consistency with other models suggested by our \textit{risk analyst} agent and to simplify the generation of the symbolic trees, outlined in \autoref{sec:sym_trees}. This means that the underlying interpretation of the parameters is model-dependent and requires human oversight in their application. We note that these parameters are then calibrated and the model discovery loop can suggest any number of parameters. 

First, the CEV model is described by  
\begin{equation}
\diff S_{t}=a S_{t}\diff t+b S_{t}^{c }\diff W_{t}
\label{eq:CEV}
\end{equation}
in which $S_t$ is the stochastic variable, representing the price of the security over time, $a$ is the strength of the drift term and $b$ is the elasticity parameter, denoting the strength between volatility and the asset price and $dW_t$ denotes Brownian noise. This SDE is widely used in modelling asset price dynamics, especially when the volatility is thought to depend non-linearly on the asset price \cite{beckers1980constant}. 

The CIR model is described by
\begin{equation}
\diff S_{t}=a(b-S_{t})\diff t+c {\sqrt {S_{t}}}\diff W_{t}
\label{eq:CIR}
\end{equation}
in which $S_t$ is the stochastic variable, and $a, b, c$, are model parameters. We note that the CIR model is typically used to model interest rates, where $S_t$ would be replaced by, e.g. $r_t$ to denote the interest rate and $a,b, c$ denotes the mean-reversion speed, long-term mean, and volatility strength, respectively \cite{overbeck1997estimation}. Here, we use $S_t$ for consistency with the other tested SDEs. We include this SDE as it is relatively simple and widely used, though we consider only equity price paths in later sections. 

Finally, our jump-diffusion (JD) model is described by
\begin{equation}
\diff S_{t} = a S_{t}\diff t + b S_{t} \diff W_{t} + S_t \diff J_{t}
\label{eq:jump_diffusion}
\end{equation}
in which $a$ is the expected rate of return, $b$ is volatility, which we assume to be constant, and $\diff J_t$ is a jump process, typically modelled as a Poisson process, $\diff N_t$,  with intensity parameter $\lambda$ \cite{cheang2012modern}. We note that the jump term is often further expanded as $(Y - 1)\diff N_t$, for random variable $Y$, representing the (log-normal) intensity of jumps. Here, we specify the ground-truth target equation as a simpler jump diffusion term but allow the model discovery process to identify models of increasing complexity. 

\begin{figure*}[!t]
\centering

\begin{minipage}[t]{0.48\textwidth}
  \centering
  \includegraphics[width=\linewidth]{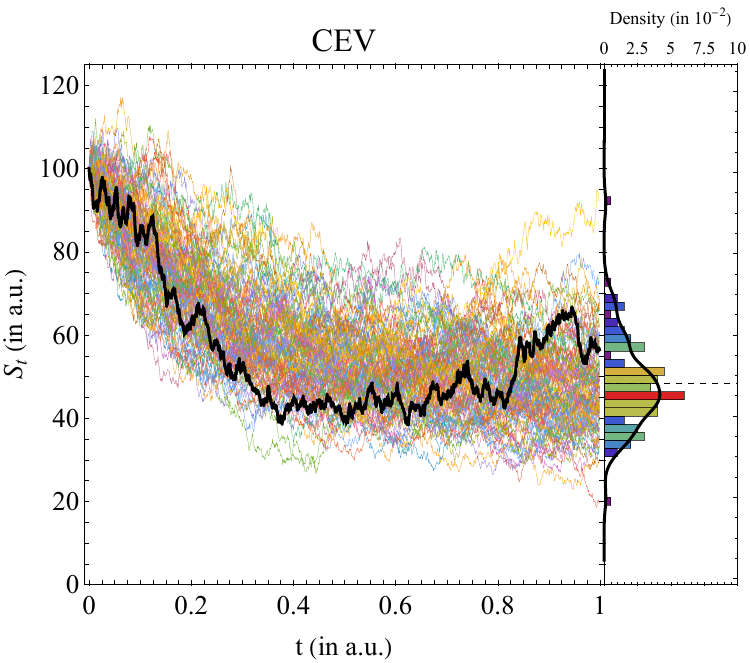}
  \label{fig:sub1}
\end{minipage}
\hfill
\begin{minipage}[t]{0.48\textwidth}
  \centering
  \includegraphics[width=\linewidth]{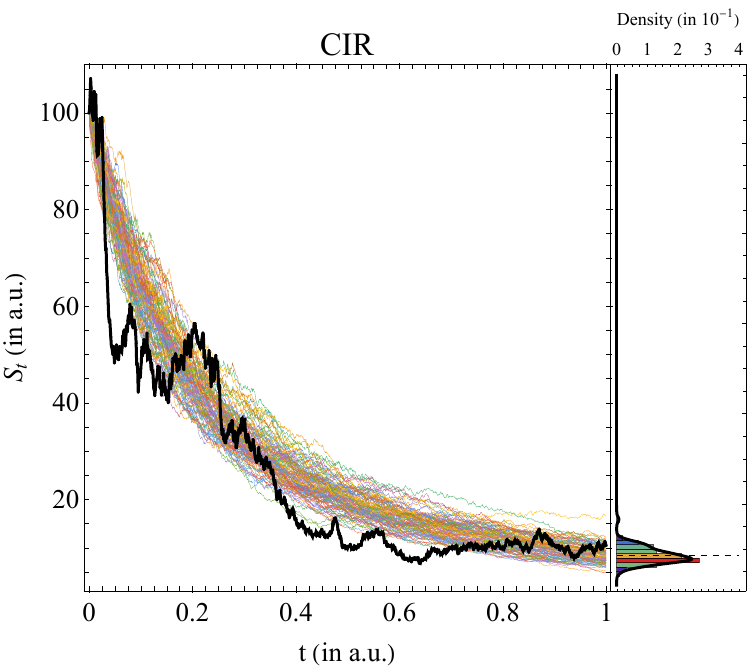}
  \label{fig:sub2}
\end{minipage}

\vspace{1em}

\begin{minipage}[c]{0.48\textwidth}
  \centering
  \includegraphics[width=\linewidth]{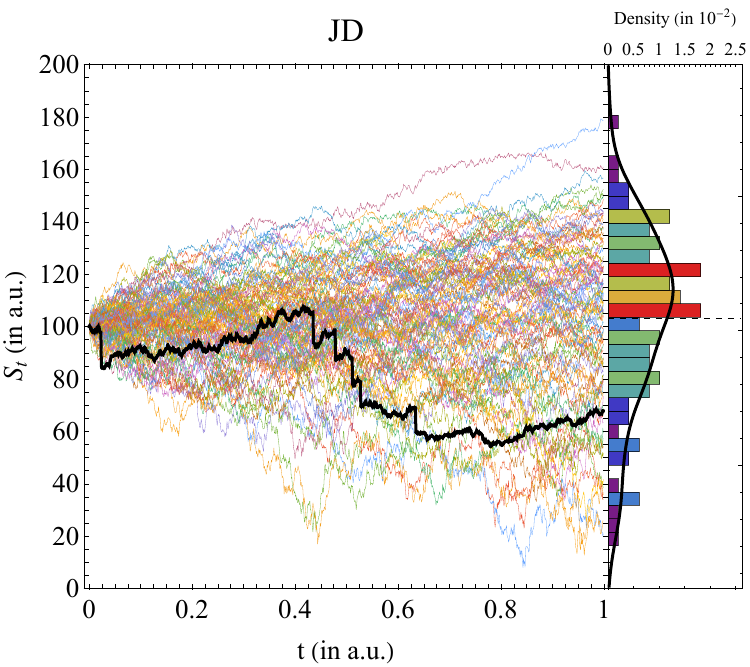}
  \label{fig:sub3}
\end{minipage}
\hfill
\begin{minipage}[c]{0.48\textwidth}
  \centering
  \vspace{-1.5em} 
  \begin{flushleft}
  \captionof{figure}{\textit{Model discovery loop examples}. To examine how effective the model discovery loop is at identifying suitable candidate models of the SDE we examine the time series of the ground truth (black) and an array of Monte Carlo simulations of the candidate model. All time, $t$, and values, $S_t$ are in arbitrary units (a.u.).\\[0.5em]
    CEV Best model: $\mathrm{d} S = a\sqrt{S}(b-S) \mathrm{d} t + c\frac{S}{1+d S} \mathrm{d} W$,\\[0.5em]
    CIR Best model: $\mathrm{d} S = a(\sqrt{S} - S) \mathrm{d} t + bS^{3/4} \mathrm{d} W$,\\[0.5em]
    JD Best model: $\mathrm{d} S = a\log(S) \mathrm{d} t + b\,\mathrm{d} W + c\,\mathrm{d} J$\\[0.5em]
    See, \autoref{eq:CEV}, \autoref{eq:CIR}, and \autoref{eq:jump_diffusion} for ground-truth symbolic equations.
  }
  \label{fig:all_eqs}
  \end{flushleft}
\end{minipage}
\end{figure*}

We find that the majority of LLMs are able to identify candidate models that can recreate the target path generated by each of the above SDEs. This is shown in \autoref{tab:equations_performance}, in which most models have a reasonably low loss. The loss is calculated from the calibration process, given in \autoref{eq:var_loss_meas}, representing path similarity. However, across all target SDEs, we find that \texttt{Sonnet 3.7} (Anthropic) consistently outperforms other LLMs, typically by several orders of magnitude. The code generated by all LLMs was typically able to run and simulate the observed time series. Additionally, we observed that different LLMs would sometimes choose and implement the same SDE with the same program, but these implementations would result in a different end result. In these cases, the differences in generating synthetic price paths is due to the parameter initialisation rather than the suggested SDE. This implies that \texttt{Sonnet 3.7} (the best model) is adept at not only program implementation but also initialising a stochastic model with a reasonable first estimate for the initial values. 

While \texttt{Sonnet 3.7} was extremely good at both implementing and calibrating SDEs, when compared to other LLMs, the proposed models were often slightly different to the actual target SDE. This is shown in \autoref{fig:all_eqs}. To quantify this, we use the symbolic similarity, described in \autoref{sec:model_disc}, which uses the Weisfeiler-Lehman kernel to compare among symbolic trees, representing the underlying SDE. While \texttt{Sonnet 3.7} often suggests similar models to the ground-truth SDEs, $m^*$, we found that the most similar models were typically suggested by OpenAI series such as \texttt{o3-mini} and \texttt{4o-mini}. Additionally, \texttt{llama-3.3} (Meta) achieved very good model similarity for the first equation (CIR) but rarely suggested other models and often failed to correctly implement these models. In fact, when considering the diversity of the models suggested, $K_\mathrm{diversity}$, given by \autoref{eq:diversity},  \texttt{o1-mini} (OpenAI) was considerably more creative and explorative in suggesting different model terms. These results highlight that LLMs will act in non-obvious and heterogenous ways within the same agentic framework, even when the temperature is set equally (which we set to be 0.2 where possible in this research). This strengthens the need for robust testing when integrating LLMs into analysis workflows. 

Finally, we note that we tested a mix of both reasoning and standard LLMs. In general, we find that reasoning LLMs, as is now well documented, typically perform better than their non-reasoning counterparts \cite{ferrag2025llm, ke2025survey}. This is true for \texttt{Sonnet 3.7} compared to \texttt{Sonnet 3.5v2} as well as \texttt{o3-mini} and \texttt{o1-mini} compared to \texttt{4o-mini}. In general, we find that LLMs that have reasoning capabilities have a `reasoning bonus' when considering the combined score, which is a weighted combination of loss, LLM score and the similarity to the target equation. However, despite this `reasoning bonus', we still find that non-reasoning LLMs can be highly performant such as the \texttt{Sonnet 3.5v2}, which was consistently within the top three best LLMs for these experiments. 

\subsection{Trading based on Stochastic Models} \label{sec:trading_results}
In this section we investigate the ability of LLMs to perform model discovery within realistic trading environments. We test the ability of each LLM to make trading decisions for the following equity symbols: \textit{AAPL} (Apple), \textit{NVDA} (NVIDIA), \textit{F} (Ford) and \textit{MSFT} (Microsoft). These companies were chosen due to the availability of news sources and for the interesting price dynamics over the chosen period. For example, NVIDIA experienced a significant fall in market price during this period, Microsoft and Ford had price paths with a similar start and end price, requiring a more active trading strategy in order to make a significant profit, and Apple had strong positive price movements.

For each asset, we repeat the model discovery loop fifteen times to find the most performant SDE model. This SDE is then used to generate the risk and trend metrics as described in \autoref{sec:trend_metrics}, and these are then passed to an LLM (along with the news context) to make informed trading decisions. We evaluate this trading ability from 17\textsuperscript{th} September 2024 to 24\textsuperscript{th} April 2025. This was chosen to ensure that the trading period is outside of the LLMs training data \cite{wang2024llmknowledge}. We then repeat the model discovery process for the same LLMs as described in \autoref{sec:model_disc_results}. 

Each LLM is prompted to make a trading decision (buy, hold, or sell) given either solely news context from the last five days, \textit{context}$^\mathrm{N}$, or news context together with the risk and trend metrics, \textit{context}$^\mathrm{N}_\mathrm{M}$, as generated by the model discovery loop. Trading decisions are then made on a daily basis over the trading period. Further details on the \textit{trader} agent are given in \autoref{sec:traders}. The trading strategy is then evaluated using three metrics:
\begin{enumerate}
\item Profit and Loss (PnL) (in \$)
\begin{equation}
\mathrm{PnL} = \text{Final Portfolio Value} - \text{Initial Portfolio Value}.
\label{eq:PnL}
\end{equation}

\item Sharpe ratio (SR)
\begin{equation}
SR = \frac{R_p-R_f}{\sigma_p}
\label{eq:rsk_adj_roi}
\end{equation}
in which $R_p$ is the portfolio return, $R_f$ is the risk free return and $\sigma_p$ is the volatility of portfolio returns.

\item  Maximum Drawdown (MDD) is given by \autoref{eq:mdd} and is a relative amount.
\end{enumerate} 
The initial portfolio value is \$1000 and we use the three-month US treasury bill as our risk-free rate. For all symbols, we also compare our strategy against a simple buy-and-hold strategy.

\begin{table*}[!t]
\centering
\footnotesize
\begin{tabular}{|l|c|c|c|c|c|c|c||c|}
\hline
 & R1 & Sonnet 3.7 & Sonnet 3.5v2 & 4o-mini & o1-mini & o3-mini & Llama 3.3 &  B\&H \\
\hline
\multicolumn{9}{|c|}{\textit{AAPL} (Apple)} \\
\hline
PnL  (N)           & 215 & 61 & 100 & 347 & 166 & 293 & \textbf{431} & \underline{372} \\
SR  (N)            & 2.39 & 0.40 & 0.84 & 3.70 & 1.50 & 2.67 & \textbf{4.56} & 3.53 \\
MDD   (N)          & 0.12 & 0.15 & 0.14 & \textbf{0.07} & 0.10 & 0.12 & \textbf{0.07} & 0.12 \\
\hdashline
PnL (N+M)   & 230 & 337 & 75 & 328 & -- & \underline{384} & -- & 372 \\
SR (N+M)    & 2.49 & 3.27 & 0.60 & 3.05 & -- & \underline{3.87} & -- & 3.53 \\
MDD (N+M)   & 0.10 & \underline{0.08} & 0.14 & 0.12 & -- & \textbf{0.07} & -- & 0.12 \\
\hline
\multicolumn{9}{|c|}{\textit{NVDA} (NVIDIA)} \\
\hline
PnL  (N)           & 450 & 448 & 234 & 521 & 314 & 488 & 452 & 593 \\
SR  (N)            & 1.98 & 2.04 & 1.04 & 2.37 & 1.37 & 2.19 & 1.98 & 2.79 \\
MDD  (N)           & 0.27 & 0.27 & 0.25 & 0.27 & 0.31 & 0.28 & 0.28 & 0.27 \\
\hdashline
PnL (N+M)   & 32 & \underline{747} & -- & 323 & -34 & 602 & \textbf{877} & 593 \\
SR (N+M)    & 0.05 & \underline{4.03} & -- & 1.36 & -0.24 & 2.86 & \textbf{5.19} & 2.79 \\
MDD (N+M)   & 0.39 & \textbf{0.23} & -- & 0.28 & 0.38 & 0.27 & \underline{0.24} & 0.27 \\
\hline
\multicolumn{9}{|c|}{\textit{MSFT} (Microsoft)} \\
\hline
PnL   (N)          & 19 & 60 & 17 & 22 & 18 & 53 & 22 & 29 \\
SR (N)              & -0.03 & 0.46 & -0.04 & 0.02 & -0.04 & 0.37 & 0.01 & 0.09 \\
MDD   (N)             & 0.15 & 0.11 & 0.16 & 0.12 & 0.12 & 0.15 & 0.15 & 0.15 \\
\hdashline
PnL (N+M)   & \underline{106} & \textbf{163} & \underline{106} & -41 & 6 & -- & -- & 29 \\
SR (N+M)    & \underline{1.17} & \textbf{1.87} & 1.12 & -0.73 & -0.24 & -- & -- & 0.09 \\
MDD (N+M)   & \textbf{0.07} & \textbf{0.07} & \textbf{0.07} & 0.15 & \underline{0.08} & -- & -- & 0.15 \\
\hline
\multicolumn{9}{|c|}{ \textit{F} (Ford)} \\
\hline
PnL  (N)              & 6 & 5 & 78 & 63 & -46 & -140 & -130 & -146 \\
SR    (N)             & -0.13 & -2.26 & 0.92 & 0.67 & -0.40 & -0.83 & -0.94 & -0.84 \\
MDD    (N)            & 0.12 & \textbf{0.00} & 0.06 & 0.05 & 0.22 & 0.33 & 0.24 & 0.33 \\
\hdashline
PnL (N+M)   & -- & -306 & \underline{179} & -145 & \textbf{180} & -44 & -72 & -146 \\
SR (N+M)    & -- & -1.86 & \textbf{2.96} & -0.98 & \underline{1.82} & -0.37 & -0.57 & -0.84 \\
MDD (N+M)   & -- & 0.33 & \underline{0.03} & 0.29 & 0.15 & 0.33 & 0.32 & 0.33 \\
\hline
\end{tabular}
\caption{\textit{Trading results by symbol}. Profit  and loss (PnL) is based on \$1000 initial portfolio value, with units of dollars, while maximum drawdown (MDD) is given in relative terms. Sharpe ratio (SR) is unitless. N denotes news context (\textit{context}$^\mathrm{N}$) and N+M denotes news context with risk and trend metrics (\textit{context}$^\mathrm{N}_\mathrm{M}$). B\&H corresponds to a buy-and-hold strategy. Note that for PnL and SR, larger values equate to a higher performance, and for MDD a higher performance is depicted by a result closer to zero. \textbf{Bold}=best, \underline{underlined}=second best and the dash (--) indicates that the model failed to complete the model discovery process. }
\label{tab:trading_results}
\end{table*}

\autoref{tab:trading_results} shows our results for the tested LLMs. We see an average SR of 0.88 when trading with just the news context. When including the model discovery loop to generate the risk and trend metrics, this average SR increases to 1.40. This shows that on average the model discovery loop enhances the SR of the trading strategies 37\%.

This is especially true for \texttt{Llama 3.3} and \texttt{o3-mini}. For \texttt{Sonnet 3.7} and \texttt{Sonnet 3.5v2} the performance when comparing \textit{context}$^\mathrm{N}$ and \textit{context}$^\mathrm{N}_\mathrm{M}$ is inconsistent. When the \textit{context}$^\mathrm{N}_\mathrm{M}$ performance is better, however, this improvement is significant (roughly more than twice the PnL). As an example, when considering Microsoft, the PnL for \texttt{Sonnet 3.7} increases from 60 to 163 with the additional metrics context, similarly with \texttt{Sonnet 3.5v2} the PnL increases from 17 to 106, showing a large improvement. 

For \texttt{4o-mini} and \texttt{o1-mini} the performance worsens with \textit{context}$^\mathrm{N}_\mathrm{M}$ compared to \textit{context}$^\mathrm{N}$. These LLMs performed poorly at the model discovery step as shown in \autoref{tab:equations_performance}. This suggests that if the LLM is working with a poor model of the financial time series it will likely not be effective at producing profitable trading strategies. In addition to selecting suboptimal models, poor performance during model discovery is often caused by SDE models that fail to compile due to incorrect coding strategies. Rather than introducing complicated fallback solutions or attempting per-LLM prompt engineering, we tried to agentically debug each of these LLMs three times before marking this SDE as a failure. A loop failure is recorded when all fifteen SDE models selected during the model discovery process fail, we denote this as a dash (--), we show this in \autoref{tab:trading_results}. We note that almost all LLMs showed at least one model failure during the model discovery process except for \texttt{Sonnet 3.7} and \texttt{4o-mini}, which we believe is due to their superior coding capabilities. The LLM with the highest failure rate is \texttt{Llama 3.3} as we discuss below. 

\begin{figure*}[!p]
    \centering
    \begin{subfigure}{0.48\textwidth}
        \centering
        \includegraphics[width=\linewidth]{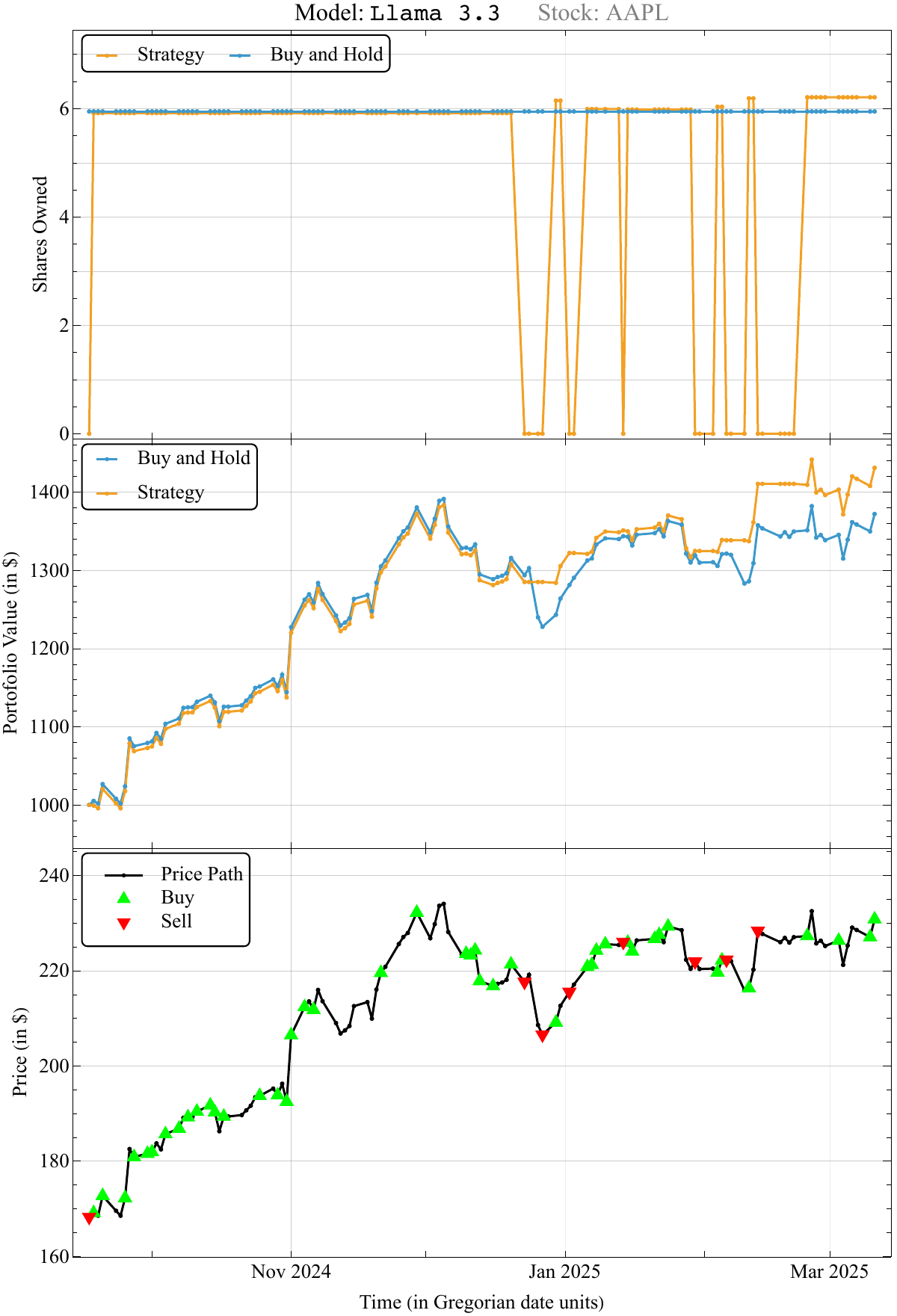}
        \label{fig:AAPL_trades}
    \end{subfigure}
    \begin{subfigure}{0.48\textwidth}
        \centering
        \includegraphics[width=\linewidth]{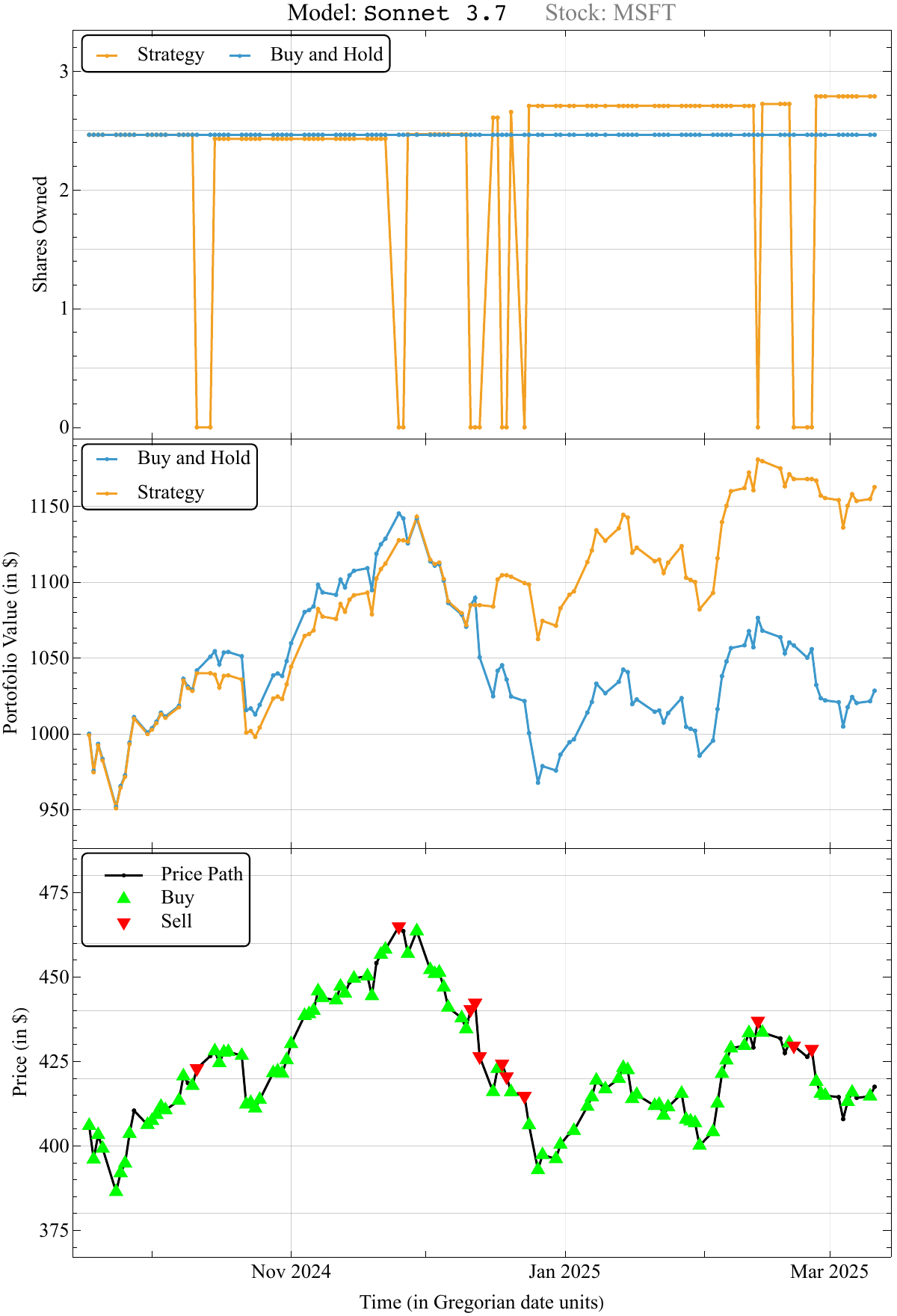}
        \label{fig:MSFT_trades}
    \end{subfigure}
    \\
    \begin{subfigure}{0.48\textwidth}
        \centering
        \includegraphics[width=\linewidth]{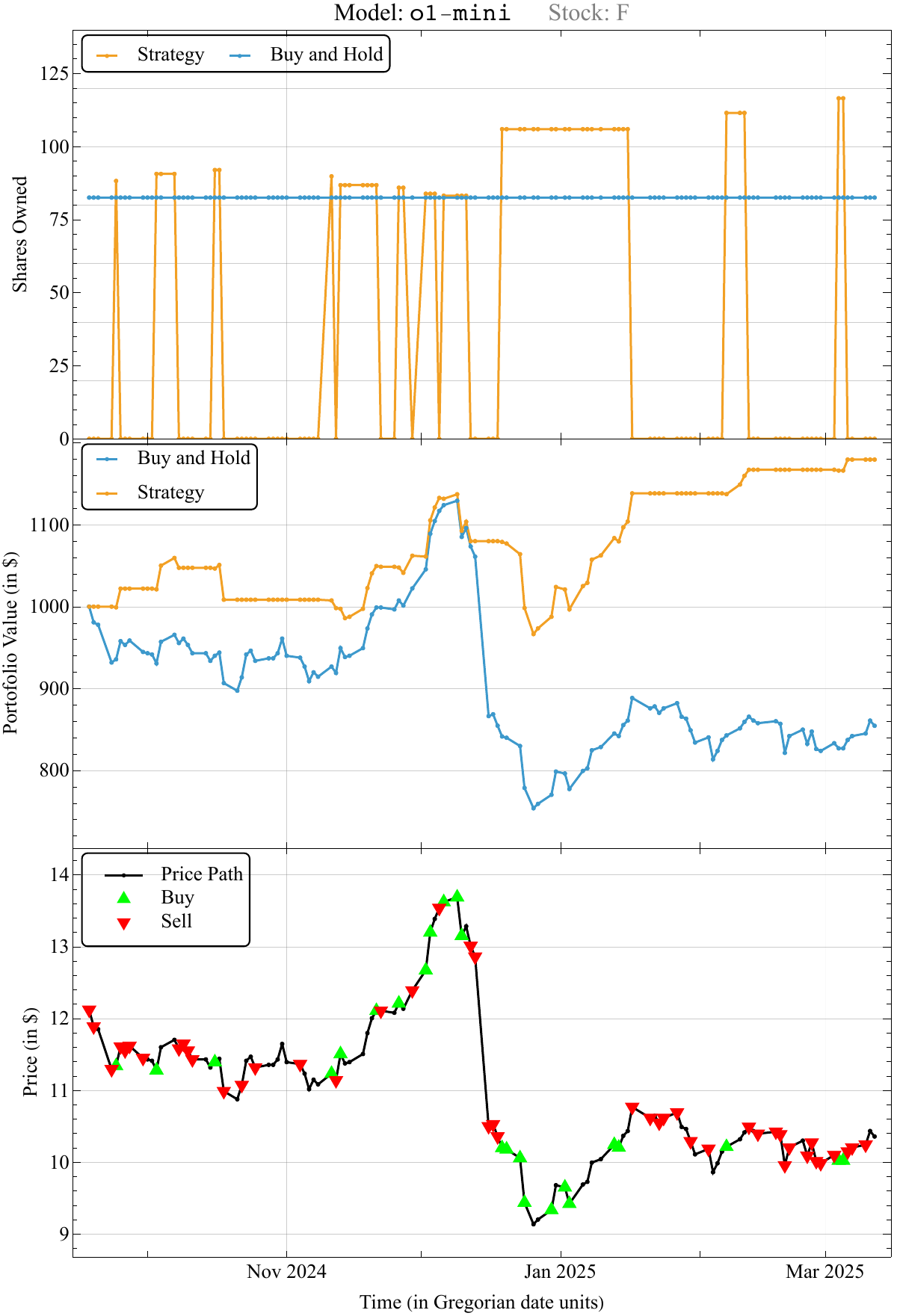}
        \label{fig:F_trades}
    \end{subfigure}
    \begin{subfigure}{0.48\textwidth}
        \centering
        \includegraphics[width=\linewidth]{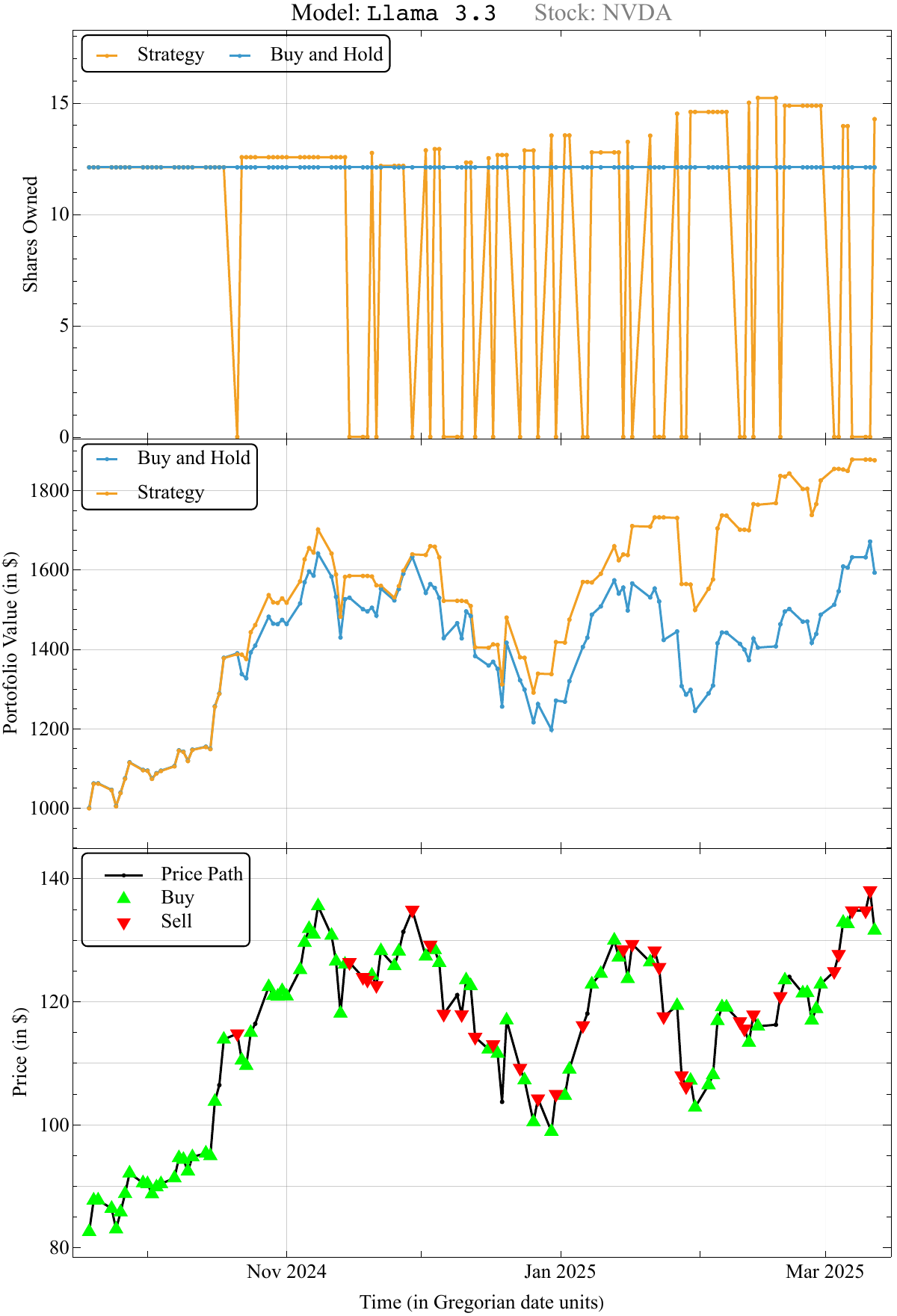}
        \label{fig:NVDA_trades}
    \end{subfigure}
    
    \caption{\textit{Examples of trading results for each of the symbols considered}. Backtesting results from agentic trading strategies shown in orange for the period covering 17\textsuperscript{th} September 2024 to 24\textsuperscript{th} April 2025. We show results for \texttt{Llama 3.3} trading on \textit{AAPL}, \texttt{Sonnet 3.7} trading on \textit{MSFT}, \texttt{o1-mini} trading on \textit{F},  \texttt{Llama 3.3} trading on \textit{NVDA}. These represent the best results for each symbol, as shown in \autoref{tab:trading_results}. Additionally, we show results from a buy-and-hold strategy in blue. Each panel shows trading decisions, the resulting portfolio output, and the number of shares owned during the trading period.}
    \label{fig:trading_simulation}
\end{figure*}

Interestingly, \texttt{Llama 3.3} demonstrates inconsistent capabilities at model discovery. It suggests far more complex and novel symbolic models. However, this model complexity is not complimented by sufficient coding ability. This results in a worse performance at model discovery and we regularly could not get \texttt{Llama 3.3} to complete the full trading period.

In \autoref{tab:equations_performance} we find that \texttt{Sonnet 3.7} performs the best at model discovery. However, we do not observe that this directly carries downstream to the results shown in \autoref{tab:trading_results}. Poor model discovery typically results in poor trading. Conversely, it is not always the case that an LLM that builds a good candidate SDE (as measured by a low calibration loss, for example) will go on to make profitable trades. This is may be due to a number of factors, from poor initialisation of parameter values, changes in market regimes, unexpected news events, or simply due to the prompt-construction for the \textit{trader} agent. Hence, the performance deficit between choosing an SDE model with a low calibration loss and making profitable trades will be due to our multi-component agentic framework (\autoref{fig:analyst_framework}). This framework requires LLMs to be equally good at suggesting SDE models to test, implementing these SDE models in code, analysing recent news articles and then balancing both model-informed risk statistics with news summaries to make a trading decision. This highlights the complexity inherent in building agentic trading systems. However, as discussed above, on average we find that the addition of a model discovery step improves trading results. 

Next, we compare each LLM to the baseline strategy to get a better understanding of the performance of our agentic framework. We find that, when compared to a buy-and-hold strategy, the \textit{trader} agents have mixed performance. For example, we find that for NVIDIA, no LLM is able to outperform the buy-and-hold baseline. This is likely due to the drastic increase in value of the NVIDIA stock over this period (roughly 47\% returns), making outperforming a buy-and-hold very difficult. The buy-and-hold is outperformed by one LLM for Apple (\texttt{Llama 3.3}) and three LLMs for Microsoft (\texttt{Sonnet-3.7}, \texttt{Sonnet 3.5v2} and \texttt{R1}). Apple and Microsoft both saw returns of 28\% and 3\%, respectively. This shows the efficacy of the framework to generate notable returns beyond a bullish market. For Ford, all LLMs are able to outperform the baseline. This result is significant, Ford is the only symbol to see a decrease in value over the trading period. This means that in order to beat the baseline the \textit{traders} needed to make decisions that took advantage of the short term price fluctuations in the market, something that is very difficult to do.

It is highly likely that different LLMs excel at different components within this framework, such as coding, sentiment analysis, and model discovery. Furthermore, we expect certain LLMs to perform better in non-obvious ways, such as the increased complexity of models suggested by \texttt{Llama 3.3}. Finally, the ability of LLMs appears to be linked to their reasoning ability. Models such as \texttt{R1}, \texttt{Sonnet 3.7}, and \texttt{o3-mini} are all reasoning LLMs and are often within the top two trading performances. However, this is not consistent across symbols. Future work is required to further evaluate these types of multi-agent LLM-based systems in complex tasks, as well as the application of smaller fine-tuned models.

\subsection{Context-Aware Backtesting with \texttt{Simudyne Horizon}}\label{sec:horizon_results}

In \autoref{sec:trading_results}, we demonstrate that the combination of model discovery with news analysis can lead to profitable trading strategies. Additionally, we find that the distribution of performance across LLMs is non-obvious, i.e. reasoning LLMs may not correlate to trading performance. This is linked to the LLMs ability to perform sentiment analysis on recent events and model discovery for evaluating risk statistics. We assume that the overall performance of our \textit{trader} agents will be influenced by the pre-training of the underlying LLMs, as we use a single LLM for all tasks within our agentic framework.

Assuming that all LLMs have been trained on a significant proportion of the internet up to their cut-off date, we cannot preclude the possibility of a biasing factor from their pre-training data affecting their trading performance in the equity markets. Additionally, it is not clear that we can entirely discount this even when using data explicitly outside of their training data knowledge cut-off as we have done in \autoref{sec:model_disc_results}. This is due to the LLMs potentially using previous data in their pre-training to assign probabilities to unseen events which are implicit in the LLM inference output and additional to the news summaries provided in \textit{context}$^\mathrm{N}_\mathrm{M}$. Moreover, for those LLMs that do not expose their reasoning (via their API), we cannot be sure that system prompts that are unseen have been designed to augment their responses in non-obvious ways.

To address this, we use the \texttt{Simudyne Horizon} platform to generate an alternative price path and synthetic news cycle for the S\&P 500 covering the 6 month trading period from 1\textsuperscript{st} January 2023 to 1\textsuperscript{st} July 2023. Illustratively, this includes fictional events such as the blocking of a major pharmaceutical merger, a major cyber attack on financial infrastructure, and the introduction of AI licensing by the US supreme court. Results from running the same trading agent and buy-and-hold scenario in the \texttt{Simudyne Horizon} backtesting environment are given in \autoref{tab:trading_results_horizon}. 

We find that most LLMs are able to outperform the simple buy-and-hold strategy by using news stories only, \textit{context}$^{\mathrm{N}}$. The trading results are further improved when combined with a model discovery loop, as in \autoref{sec:trading_results}. We see an average 22\% increase in SR when comparing LLMs using \textit{context}$^\mathrm{N}$ to \textit{context}$^\mathrm{N}_\mathrm{M}$. The models \texttt{4o-mini}, \texttt{o3-mini},  \texttt{Sonnet 3.5v2}, and \texttt{Sonnet 3.7} all improved with the additional risk and trend metrics, however, both \texttt{R1} and \texttt{o1-mini} resulted in worse performance. As in \autoref{sec:model_disc_results}, we find that \texttt{Llama 3.3} is unable to complete the model discovery due to the complexity of models that were suggested. Additionally, we find that \texttt{Sonnet 3.5v2}, and \texttt{Sonnet 3.7} and \texttt{4o-mini} resulted in profitable trading decisions across the synthetic backtest. Moreover, these results highlight that model-informed analysis improves agentic trading decisions and that these results are robust to potential pre-training biases. However, as in \autoref{sec:trading_results}, the performance of the agentic \textit{trader} agent is connected to the underlying LLM in a non-obvious way, such that reasoning LLMs do not necessarily result in improved trading performance. Our results highlight both the potential and fragility of agentic trading workflows. 

\begin{table*}
\centering
\footnotesize
\begin{tabular}{|l|c|c|c|c|c|c|c||c|}
\hline
 & Deepseek R1 & Sonnet 3.7 & Sonnet 3.5v2 & 4o-mini & o1-mini & o3-mini & LLama 3.3 &  B\&H \\
\hline
PnL (N)            & -62 & -12 & -22 & -71 & -13 & -113 & -28 & -99 \\
SR (N)             & -0.77 & -0.68 & -1.23 & -1.06 & -0.44 & -1.46 & -0.85 & -1.09 \\
MDD (N)            & 0.14 & \underline{0.06} & \textbf{0.05} & 0.10 & 0.07 & 0.16 & 0.07 & 0.16 \\
\hdashline
PnL (N+M)       & -88 & \textbf{53} & \underline{45} & 1 & -66 & -55 & -- & -99 \\
SR (N+M)        & -1.70 & \underline{0.39} & \textbf{0.47} & -0.21 & -1.07 & -1.13 & -- & -1.09 \\
MDD (N+M)       & 0.11 & \underline{0.06} & \underline{0.06} & 0.08 & 0.09 & 0.12 & -- & 0.16 \\
\hline
\end{tabular}
\caption{\textit{Trading results using the} 
\texttt{Simudyne Horizon} \textit{backtesting environment}. Profit  and loss (PnL) is based on \$1000 initial portfolio value, with units of dollars, while maximum drawdown (MDD) is given in relative terms. Sharpe ratio (SR) is unitless. N denotes news context (\textit{context}$^\mathrm{N}$) and N+M denotes news context with risk and trend metrics (\textit{context}$^\mathrm{N}_\mathrm{M}$). B\&H corresponds to the buy-and-hold strategy. Note that for PnL and SR, larger values equate to a higher performance, and for MDD a higher performance is depicted by the value closest to zero. \textbf{Bold}=best, \underline{underlined}=second best and the dash (--) indicates that the model failed to complete the model discovery process. }
\label{tab:trading_results_horizon}
\end{table*}

\section{Conclusion}

\lettrine[lines=2, lhang=0.2, loversize=0.1, findent=0.25em]{E}{\hspace{0.4em} vidence} in this research highlights that LLMs can be deployed in a novel agentic trading framework, leading to improved trading performance. Our research posed the title question on whether to trade or not to trade. We believe that this is no longer a question posed blindly to the market, nor answered solely by fundamental analysis, momentum, or gut instinct. In our research, we have shown that LLMs, embedded within an agentic framework, can participate meaningfully in the construction of explanatory models of financial time series. These models simulate, rather than simply predict, the uncertainty inherent in market dynamics. Yet perhaps the most significant contribution lies not in their performance gains, but in the semi-automated construction of an epistemic loop where discovery, simulation, and judgement co-evolve. 

By integrating symbolic model-informed context into trading agents, capable of digesting both price signals and narrative context, we have demonstrated a methodology for LLMs to make trading decisions. This method allows for model-aware risk quantification, in which the model development is partially automated by LLMs. This supports other recent works demonstrating that LLMs can be deployed to automate research and development tasks \cite{lu2024ai, gottweis2025towards}. We believe that this is the first time that this approach, LLM-based model development, has been applied to quantitative risk analysis for financial markets. Specifically, we demonstrate that our method can accurately determine the form of a target SDE, as outlined in \autoref{sec:model_disc_results}. 

Our model discovery loop was used to estimate the market risk to enhance trading strategies. Here, we observed that model-informed risk and trend metrics improved trading decisions, and that this represented a considerable improvement against a simple buy-and-hold strategy. However, we believe future work should consider the implications of such systems in the context of black swan events. In this research, we have found a distinct difference in the capabilities of the tested LLMs and expect that the specific skill-set of individual LLMs may be more effectively harnessed towards a common goal by applying them to LLM-specific sub-tasks. Future work will investigate the advantage of using different LLMs for individual tasks within a comprehensive agentic trading framework. 

This study represents a first step towards the introduction of automated LLM-based reasoning agents in financial analysis and trading systems. We allow for informed and explainable decisions through the construction of interpretable programs representing simulators of financial time series. This provides clarity on decision-making. Most importantly, we highlight a considerable opportunity for profitable strategies. Yet, at the same time, we observe plausible risks inherent to the current generation of LLMs if used in a financial multi-agent framework. This is demonstrated by the LLMs inconsistent and varied performance in various steps of our agentic workflow. We expect that both opportunities and risks will compound in the future. As agentic systems mature, the question will not be to trade, but how to trade wisely --- being mindful that the future is not predicted, but emerges through the recursive intelligence of models that learn to reason, simulate, and decide.

\section{Acknowledgements}
This work was generously supported by compute from AWS and NVIDIA. In particular, we thank Ilan Gleiser and the HPC and GenAI teams at AWS. We also thank Mark Atkin and Richard Harmon for their helpful comments on our manuscript.

\addtocategory{nonhuman}{opus2025illusion}

\printbibliography[notcategory=nonhuman,title={References}]

\printbibliography[category=nonhuman,title={LLM Co-authored References}]

@article{opus2025illusion,
  title={The Illusion of the Illusion of Thinking},
  author={Opus, C and Lawsen, A},
  journal={arXiv preprint arXiv:2:2506.09250},
  year={2025}
}

@article{yao2024tau,
  title={tau-bench: A Benchmark for Tool-Agent-User Interaction in Real-World Domains},
  author={Yao, Shunyu and Shinn, Noah and Razavi, Pedram and Narasimhan, Karthik},
  journal={arXiv preprint arXiv:2406.12045},
  year={2024}
}

@article{shojaee2025illusion,
  title={The illusion of thinking: Understanding the strengths and limitations of reasoning models via the lens of problem complexity},
  author={Shojaee, Parshin and Mirzadeh, Iman and Alizadeh, Keivan and Horton, Maxwell and Bengio, Samy and Farajtabar, Mehrdad},
  journal={arXiv preprint arXiv:2506.06941},
  year={2025}
}

@article{ferrag2025llm,
  title={From llm reasoning to autonomous ai agents: A comprehensive review},
  author={Ferrag, Mohamed Amine and Tihanyi, Norbert and Debbah, Merouane},
  journal={arXiv preprint arXiv:2504.19678},
  year={2025}
}

@misc{embrechts2011quantitative,
  title={Quantitative Risk Management.},
  author={Embrechts, Paul and Frey, Rdiger and McNeil, Alexander},
  year={2011}
}

@book{hollander2013nonparametric,
  title={Nonparametric statistical methods},
  author={Hollander, Myles and Wolfe, Douglas A and Chicken, Eric},
  year={2013},
  publisher={John Wiley \& Sons}
}

@book{villani2008optimal,
  title={Optimal transport: old and new},
  author={Villani, C{\'e}dric and others},
  volume={338},
  year={2008},
  publisher={Springer}
}

@book{hyndman2018forecasting,
  title={Forecasting: principles and practice},
  author={Hyndman, Rob J and Athanasopoulos, George},
  year={2018},
  publisher={OTexts}
}

@inproceedings{ratliff2023revisiting,
  title={Revisiting stylized facts for modern stock markets},
  author={Ratliff-Crain, Ethan and Van Oort, Colin M and Bagrow, James and Koehler, Matthew TK and Tivnan, Brian F},
  booktitle={2023 IEEE International Conference on Big Data (BigData)},
  pages={1814--1823},
  year={2023},
  organization={IEEE}
}

@article{safari2025international,
  title={International Financial Markets Through 150 Years: Evaluating Stylized Facts},
  author={Safari, Sara A and Janisch, Maximilian and Leh{\'e}ricy, Thomas},
  journal={arXiv preprint arXiv:2504.08611},
  year={2025}
}

@book{tsay2005analysis,
  title={Analysis of financial time series},
  author={Tsay, Ruey S},
  year={2005},
  publisher={John wiley \& sons}
}

@book{cont2003financial,
  title={Financial modelling with jump processes},
  author={Cont, Rama and Tankov, Peter},
  year={2003},
  publisher={Chapman and Hall/CRC}
}

@article{goodman2024,
  title={Automated statistical model discovery with language models},
  author={Li, Michael Y and Fox, Emily B and Goodman, Noah D},
  journal={arXiv preprint arXiv:2402.17879},
  year={2024}
}

@article{lu2024ai,
  title={The ai scientist: Towards fully automated open-ended scientific discovery},
  author={Lu, Chris and Lu, Cong and Lange, Robert Tjarko and Foerster, Jakob and Clune, Jeff and Ha, David},
  journal={arXiv preprint arXiv:2408.06292},
  year={2024}
}

@article{gottweis2025towards,
  title={Towards an AI co-scientist},
  author={Gottweis, Juraj and Weng, Wei-Hung and Daryin, Alexander and Tu, Tao and Palepu, Anil and Sirkovic, Petar and Myaskovsky, Artiom and Weissenberger, Felix and Rong, Keran and Tanno, Ryutaro and others},
  journal={arXiv preprint arXiv:2502.18864},
  year={2025}
}

@article{tian2023macgyver,
  title={MacGyver: Are Large Language Models Creative Problem Solvers?},
  author={Tian, Yufei and Ravichander, Abhilasha and Qin, Lianhui and Bras, Ronan Le and Marjieh, Raja and Peng, Nanyun and Choi, Yejin and Griffiths, Thomas L and Brahman, Faeze},
  journal={arXiv preprint arXiv:2311.09682},
  year={2023}
}

@article{kumar2024human,
  title={Human Creativity in the Age of LLMs: Randomized Experiments on Divergent and Convergent Thinking},
  author={Kumar, Harsh and Vincentius, Jonathan and Jordan, Ewan and Anderson, Ashton},
  journal={arXiv preprint arXiv:2410.03703},
  year={2024}
}

@article{ke2025survey,
  title={A Survey of Frontiers in LLM Reasoning: Inference Scaling, Learning to Reason, and Agentic Systems},
  author={Ke, Zixuan and Jiao, Fangkai and Ming, Yifei and Nguyen, Xuan-Phi and Xu, Austin and Long, Do Xuan and Li, Minzhi and Qin, Chengwei and Wang, Peifeng and Savarese, Silvio and others},
  journal={arXiv preprint arXiv:2504.09037},
  year={2025}
}

@article{plaat2025agentic,
  title={Agentic Large Language Models, a survey},
  author={Plaat, Aske and van Duijn, Max and van Stein, Niki and Preuss, Mike and van der Putten, Peter and Batenburg, Kees Joost},
  journal={arXiv preprint arXiv:2503.23037},
  year={2025}
}

@article{cao2025deep,
  title={From Deep Learning to LLMs: A survey of AI in Quantitative Investment},
  author={Cao, Bokai and Wang, Saizhuo and Lin, Xinyi and Wu, Xiaojun and Zhang, Haohan and Ni, Lionel M and Guo, Jian},
  journal={arXiv preprint arXiv:2503.21422},
  year={2025}
}

@inproceedings{li2023large,
  title={Large language models in finance: A survey},
  author={Li, Yinheng and Wang, Shaofei and Ding, Han and Chen, Hang},
  booktitle={Proceedings of the fourth ACM international conference on AI in finance},
  pages={374--382},
  year={2023}
}

@article{nie2024survey,
  title={A survey of large language models for financial applications: Progress, prospects and challenges},
  author={Nie, Yuqi and Kong, Yaxuan and Dong, Xiaowen and Mulvey, John M and Poor, H Vincent and Wen, Qingsong and Zohren, Stefan},
  journal={arXiv preprint arXiv:2406.11903},
  year={2024}
}

@article{zhao2024revolutionizing,
  title={Revolutionizing finance with llms: An overview of applications and insights},
  author={Zhao, Huaqin and Liu, Zhengliang and Wu, Zihao and Li, Yiwei and Yang, Tianze and Shu, Peng and Xu, Shaochen and Dai, Haixing and Zhao, Lin and Mai, Gengchen and others},
  journal={arXiv preprint arXiv:2401.11641},
  year={2024}
}

@article{lopez2025can,
  title={Can Large Language Models Trade? Testing Financial Theories with LLM Agents in Market Simulations},
  author={Lopez-Lira, Alejandro},
  journal={arXiv preprint arXiv:2504.10789},
  year={2025}
}

@article{ding2024large,
  title={Large language model agent in financial trading: A survey},
  author={Ding, Han and Li, Yinheng and Wang, Junhao and Chen, Hang},
  journal={arXiv preprint arXiv:2408.06361},
  year={2024}
}

@inproceedings{yu2024finmem,
  title={FinMem: A performance-enhanced LLM trading agent with layered memory and character design},
  author={Yu, Yangyang and Li, Haohang and Chen, Zhi and Jiang, Yuechen and Li, Yang and Zhang, Denghui and Liu, Rong and Suchow, Jordan W and Khashanah, Khaldoun},
  booktitle={Proceedings of the AAAI Symposium Series},
  volume={3},
  number={1},
  pages={595--597},
  year={2024}
}

@incollection{cheang2012modern,
  title={A modern view on Merton's jump-diffusion model},
  author={Cheang, Gerald HL and Chiarella, Carl},
  booktitle={Stochastic Processes, Finance and Control: A Festschrift in Honor of Robert J Elliott},
  pages={217--234},
  year={2012},
  publisher={World Scientific}
}

@article{overbeck1997estimation,
  title={Estimation in the cox-ingersoll-ross model},
  author={Overbeck, Ludger and Ryden, Tobias},
  journal={Econometric Theory},
  volume={13},
  number={3},
  pages={430--461},
  year={1997},
  publisher={Cambridge University Press}
}

@article{beckers1980constant,
  title={The constant elasticity of variance model and its implications for option pricing},
  author={Beckers, Stan},
  journal={the Journal of Finance},
  volume={35},
  number={3},
  pages={661--673},
  year={1980},
  publisher={JSTOR}
}

@book{kijima2002stochastic,
  title={Stochastic processes with applications to finance},
  author={Kijima, Masaaki},
  year={2002},
  publisher={Chapman and Hall/CRC}
}

@book{rolski2009stochastic,
  title={Stochastic processes for insurance and finance},
  author={Rolski, Tomasz and Schmidli, Hanspeter and Schmidt, Volker and Teugels, Jozef L},
  year={2009},
  publisher={John Wiley \& Sons}
}

@article{shervashidze2011weisfeiler,
  title={Weisfeiler-lehman graph kernels.},
  author={Shervashidze, Nino and Schweitzer, Pascal and Van Leeuwen, Erik Jan and Mehlhorn, Kurt and Borgwardt, Karsten M},
  journal={Journal of Machine Learning Research},
  volume={12},
  number={9},
  year={2011}
}

@inproceedings{khattab2024dspy,
  title={DSPy: Compiling Declarative Language Model Calls into Self-Improving Pipelines},
  author={Khattab, Omar and Singhvi, Arnav and Maheshwari, Paridhi and Zhang, Zhiyuan and Santhanam, Keshav and Vardhamanan, Sri and Haq, Saiful and Sharma, Ashutosh and Joshi, Thomas T. and Moazam, Hanna and Miller, Heather and Zaharia, Matei and Potts, Christopher},
  journal={The Twelfth International Conference on Learning Representations},
  year={2024}
}

@article{khattab2022demonstrate,
  title={Demonstrate-Search-Predict: Composing Retrieval and Language Models for Knowledge-Intensive {NLP}},
  author={Khattab, Omar and Santhanam, Keshav and Li, Xiang Lisa and Hall, David and Liang, Percy and Potts, Christopher and Zaharia, Matei},
  journal={arXiv preprint arXiv:2212.14024},
  year={2022}
}

@article{stillman2024neurosymbolictradersassessingwisdom,  
  title={Neuro-Symbolic Traders: Assessing the Wisdom of AI Crowds in Markets},
  author={Stillman, Namid R and Baggott, Rory},
  journal={arXiv preprint arXiv:2410.14587},
  year={2024}
}

@book{mcneil2015quantitative,
  title={Quantitative risk management: concepts, techniques and tools-revised edition},
  author={McNeil, Alexander J and Frey, R{\"u}diger and Embrechts, Paul},
  year={2015},
  publisher={Princeton university press}
}

@book{coles2001introduction,
  title={An introduction to statistical modeling of extreme values},
  author={Coles, Stuart and Bawa, Joanna and Trenner, Lesley and Dorazio, Pat},
  volume={208},
  year={2001},
  publisher={Springer}
}

@article{wang2023alphagpthumanaiinteractivealpha,
  title={Alpha-gpt: Human-ai interactive alpha mining for quantitative investment},
  author={Wang, Saizhuo and Yuan, Hang and Zhou, Leon and Ni, Lionel M and Shum, Heung-Yeung and Guo, Jian},
  journal={arXiv preprint arXiv:2308.00016},
  year={2023}
}

@article{lee2024quantifyingqualitativeinsightsleveraging,
  title={Quantifying Qualitative Insights: Leveraging LLMs to Market Predict},
  author={Lee, Hoyoung and Choi, Youngsoo and Kwon, Yuhee},
  journal={arXiv preprint arXiv:2411.08404},
  year={2024}
}

@article{ding2024tradexpertrevolutionizingtradingmixture,
  title={Tradexpert: Revolutionizing trading with mixture of expert llms},
  author={Ding, Qianggang and Shi, Haochen and Guo, Jiadong and Liu, Bang},
  journal={arXiv preprint arXiv:2411.00782},
  year={2024}
}

@article{li2023tradinggptmultiagentlayeredmemory,
  title={Tradinggpt: Multi-agent system with layered memory and distinct characters for enhanced financial trading performance},
  author={Li, Yang and Yu, Yangyang and Li, Haohang and Chen, Zhi and Khashanah, Khaldoun},
  journal={arXiv preprint arXiv:2309.03736},
  year={2023}
}

@article{kou2024automatestrategyfindingllm,
  title={Automate strategy finding with llm in quant investment},
  author={Kou, Zhizhuo and Yu, Holam and Luo, Junyu and Peng, Jingshu and Li, Xujia and Liu, Chengzhong and Dai, Juntao and Chen, Lei and Han, Sirui and Guo, Yike},
  journal={arXiv preprint arXiv:2409.06289},
  year={2024}
}

@article{xiao2025tradingagentsmultiagentsllmfinancial,
  title={TradingAgents: Multi-Agents LLM Financial Trading Framework},
  author={Xiao, Yijia and Sun, Edward and Luo, Di and Wang, Wei},
  journal={arXiv preprint arXiv:2412.20138},
  year={2024}
}

@article{ORLIK2024105905,
title = {Understanding uncertainty shocks and the role of black swans},
journal = {Journal of Economic Theory},
volume = {222},
pages = {105905},
year = {2024},
issn = {0022-0531},
author = {Anna Orlik and Laura Veldkamp},
}

@article{GeorgeBox1962,
 ISSN = {00401706},
 author = {G. E. P. Box and William G. Hunter},
 journal = {Technometrics},
 number = {3},
 pages = {301--318},
 publisher = {[Taylor & Francis, Ltd., American Statistical Association, American Society for Quality]},
 title = {A Useful Method for Model-Building},
 volume = {4},
 year = {1962}
}

@article{kidger2022neuraldifferentialequations,
  title={On neural differential equations},
  author={Kidger, Patrick},
  journal={arXiv preprint arXiv:2202.02435},
  year={2022}
}

@article{jiang2025ledextrainingllmsbetter,
  title={Ledex: Training LLMs to better self-debug and explain code},
  author={Jiang, Nan and Li, Xiaopeng and Wang, Shiqi and Zhou, Qiang and Hossain, Soneya B and Ray, Baishakhi and Kumar, Varun and Ma, Xiaofei and Deoras, Anoop},
  journal={Advances in Neural Information Processing Systems},
  volume={37},
  pages={35517--35543},
  year={2024}
}

@article{basyal2023textsummarizationusinglarge,
  title={Text summarization using large language models: a comparative study of mpt-7b-instruct, falcon-7b-instruct, and openai chat-gpt models},
  author={Basyal, Lochan and Sanghvi, Mihir},
  journal={arXiv preprint arXiv:2310.10449},
  year={2023}
}

@inproceedings{Chiang2023CanLL,
  title={Can Large Language Models Be an Alternative to Human Evaluations?},
  author={Cheng-Han Chiang and Hung-yi Lee},
  booktitle={Annual Meeting of the Association for Computational Linguistics},
  year={2023}
}

@article{faldor2025omniepicopenendednessmodelshuman,
  title={OMNI-EPIC: Open-endedness via Models of human Notions of Interestingness with Environments Programmed in Code},
  author={Faldor, Maxence and Zhang, Jenny and Cully, Antoine and Clune, Jeff},
  journal={arXiv preprint arXiv:2405.15568},
  year={2024}
}

@article{openai2024openaio1card,
  title={Openai o1 system card},
  author={Jaech, Aaron and Kalai, Adam and Lerer, Adam and Richardson, Adam and El-Kishky, Ahmed and Low, Aiden and Helyar, Alec and Madry, Aleksander and Beutel, Alex and Carney, Alex and others},
  journal={arXiv preprint arXiv:2412.16720},
  year={2024}
}

@article{deepseekai2025deepseekr1incentivizingreasoningcapability,
  title={Deepseek-r1: Incentivizing reasoning capability in llms via reinforcement learning},
  author={Guo, Daya and Yang, Dejian and Zhang, Haowei and Song, Junxiao and Zhang, Ruoyu and Xu, Runxin and Zhu, Qihao and Ma, Shirong and Wang, Peiyi and Bi, Xiao and others},
  journal={arXiv preprint arXiv:2501.12948},
  year={2025}
}

@article{chio2022comparativestudymacdbasetrading,
  title={A comparative study of the MACD-base trading strategies: evidence from the US stock market},
  author={Chio, Pat Tong},
  journal={arXiv preprint arXiv:2206.12282},
  year={2022}
}

@misc{wang2024llmknowledge,
  author = {Wang, Hao},
  title = {LLM Knowledge Cutoff Dates},
  year = {2024},
  url = {https://github.com/HaoooWang/llm-knowledge-cutoff-dates},
  note = {GitHub repository}
}

@article{Engle2006MeasuringAM,
  title={Measuring and Modeling Execution Cost and Risk},
  author={Robert F. Engle and R. Ferstenberg and Jeffrey R. Russell},
  journal={IO: Firm Structure},
  year={2006},
}

@article{hughes2024openendednessessentialartificialsuperhuman,
  title={Open-endedness is essential for artificial superhuman intelligence},
  author={Hughes, Edward and Dennis, Michael and Parker-Holder, Jack and Behbahani, Feryal and Mavalankar, Aditi and Shi, Yuge and Schaul, Tom and Rocktaschel, Tim},
  journal={arXiv preprint arXiv:2406.04268},
  year={2024}
}

@article{wang2024applicationnaturallanguageprocessing,
  title={Application of natural language processing in financial risk detection},
  author={Wang, Liyang and Cheng, Yu and Xiang, Ao and Zhang, Jingyu and Yang, Haowei},
  journal={arXiv preprint arXiv:2406.09765},
  year={2024}
}

@article{nie2024surveylargelanguagemodels,
  title={A survey of large language models for financial applications: Progress, prospects and challenges},
  author={Nie, Yuqi and Kong, Yaxuan and Dong, Xiaowen and Mulvey, John M and Poor, H Vincent and Wen, Qingsong and Zohren, Stefan},
  journal={arXiv preprint arXiv:2406.11903},
  year={2024}
}

@article{Zhang_2013,
   title={Stochastic modeling and fair valuation of drawdown insurance},
   volume={53},
   ISSN={0167-6687},
   number={3},
   journal={Insurance: Mathematics and Economics},
   publisher={Elsevier BV},
   author={Zhang, Hongzhong and Leung, Tim and Hadjiliadis, Olympia},
   year={2013},
   month=nov, pages={840–850} }

@article{Rockafellar2000OptimizationOC,
  title={Optimization of conditional value-at risk},
  author={R. Tyrrell Rockafellar and Stanislav Uryasev},
  journal={Journal of Risk},
  year={2000},
  volume={3},
  pages={21-41},
}

@article{ding2024longropeextendingllmcontext,
  title={Longrope: Extending llm context window beyond 2 million tokens},
  author={Ding, Yiran and Zhang, Li Lyna and Zhang, Chengruidong and Xu, Yuanyuan and Shang, Ning and Xu, Jiahang and Yang, Fan and Yang, Mao},
  journal={arXiv preprint arXiv:2402.13753},
  year={2024}
}

\newpage
\appendix
\section*{Appendix}

\section{Example Agent Output} \label{app:prompt}

\begin{tcolorbox}[
  enhanced,
  colback=green!5!white,
  colframe=green!75!black,
  arc=4mm,
  title=Summarising Agent,
  fonttitle=\bfseries\large,
  coltitle=white,
  colbacktitle=green!75!black,
  attach boxed title to top left={yshift=-2mm, xshift=10mm}
]
This is a non-linear SDE with interesting properties:
Looking at the model code, we can see:
\begin{itemize}
\item The drift term is a*sin(S/b) where S is the state variable
\item The diffusion term is c*S where c is a constant
\item This forms a stochastic differential equation (SDE)
\item The equation has a nonlinear sinusoidal drift term which is interesting
\item The diffusion term is proportional to S which is standard 
\end{itemize}
\end{tcolorbox}
\begin{tcolorbox}[
  enhanced,
  colback=red!5!white,
  colframe=red!75!black,
  arc=4mm,
  title=Scorer Agent,
  fonttitle=\bfseries\large,
  coltitle=white,
  colbacktitle=red!75!black,
  attach boxed title to top left={yshift=-2mm, xshift=10mm}
]
Let's analyze this model in two aspects:
\begin{enumerate}
\item 
    {
    Novelty (50\% of score): 
    \begin{itemize}
        \item The model uses a sine function which is a novel approach for SDEs
        \item The structure dS = a*sin(S/b)dt + cSdW is relatively simple but introduces non-linear behavior
        \item  The sine function could capture cyclical patterns in the data
        \item  Score for novelty: 35/50 (good originality but maintains reasonable simplicity)
        \end{itemize}
    }
\item{Metrics Accuracy (50\% of score):
        \begin{itemize}
        \item Delta metrics show good alignment in mean (0.001) and jump intensity (0.005)
        \item  Volatility is reasonably close (delta of 0.278)
        \item  ACF values are fairly close to target (deltas around 0.02-0.04)
        \item  Kurtosis and skewness show moderate differences
        \item  The range and distribution metrics (min, max, mean) show some deviation
        \item  Kolmogorov-Smirnov test value of 0.276 indicates moderate fit
        \item  Score for metrics: 32/50
        \end{itemize}
    }
\end{enumerate}

Total score calculation:
35 (novelty) + 32 (metrics) = 67/100

\end{tcolorbox}

\begin{tcolorbox}[
  enhanced,
  colback=orange!5!white,
  colframe=orange!75!black,
  arc=4mm,
  title=Model Suggestor Agent,
  fonttitle=\bfseries\large,
  coltitle=white,
  colbacktitle=orange!75!black,
  attach boxed title to top left={yshift=-2mm, xshift=10mm}
  ]
\begin{itemize}
    \item $dS = a(b-S) dt + c\sqrt{S} dW$
    \item $dS = aS(b-\ln(S)) dt + cS dW$
    \item $dS = aS^b dt + c dW$
    \item $dS = kS(1-S) dt + mS^{1/2} dW$
    \item $dS = k(S/(1+S^2)) dt + mS dW$
    \item $dS = a\tanh(bS) dt + cS dW$
    \item $dS = k(1-\exp(-mS)) dt + p\sqrt{S} dW$
    \item $dS = a/(1+\exp(-bS)) dt + c(1+S) dW$
    \item $dS = k\sin(mS) dt + p\cos(S) dW$
    \item $dS = aS(1-\ln(S/b)) dt + cS^{1/3} dW$
    \item $dS = kS/(1+S^3) dt + m\exp(-pS) dW$
    \item $dS = a\sin(S/b) dt + cS dW$
    \item $dS = kS/(1+S^2) dt + mS/(1+S) dW$
    \item $dS = k(1-S^{1/3})dt + mS^{1/3}dW$
    \item $dS = aS\log(1+bS)dt + c/(1+S)dW$
    \item $dS = k\arctan(mS)dt + pS dW$
    \item $dS = a\exp(-bS)dt + c\cdot S/(1+S^2) dW$
    \item $dS = a\sin(bS)/(1+S^2)dt + cS dW$
    \item $dS = k\cdot S\log(1+mS)dt + p\cdot S\log(1/(1+S))dW$
    \item $dS = k\log(1+S)/(1+S^2)dt + m\cdot S/(1+\log(S))dW$
    \item $dS = a\cdot S^{1/2}(1-\exp(-bS))dt + c(1+S^{1/3})dW$
\end{itemize}
\end{tcolorbox}

\begin{tcolorbox}[
  enhanced,
  colback=blue!5!white,
  colframe=blue!75!black,
  arc=4mm,
  title=News Gatherer Agent,
  fonttitle=\bfseries\large,
  coltitle=white,
  colbacktitle=blue!75!black,
  attach boxed title to top right={{xshift=-10mm, yshift=-2mm}}
]
Based on the news articles, there's significant focus on DeepSeek, a Chinese AI startup that's causing market disruption in the AI sector. The news suggests this is affecting various tech companies, particularly those involved in AI and semiconductors. To make a comprehensive analysis of NVDA's current state, we should look at:
\begin{enumerate}
    \item Direct competitors in the GPU/AI chip space (AMD, INTC)
    \item Key semiconductor manufacturing partners (TSM)
    \item Major AI technology companies affected by DeepSeek developments (MSFT, GOOGL)
    \item Companies involved in AI infrastructure (ORCL)
\end{enumerate}

These companies will provide a complete picture of the competitive landscape, supply chain, and broader AI market dynamics affecting NVDA.
\end{tcolorbox}

\begin{tcolorbox}[
  enhanced,
  colback=purple!5!white,
  colframe=purple!75!black,
  arc=4mm,
  title=News Analyst Agent,
  fonttitle=\bfseries\large,
  coltitle=white,
  colbacktitle=purple!75!black,
  attach boxed title to top right={{xshift=-10mm, yshift=-2mm}}
]
The emergence of DeepSeek and other competitive threats suggests NVIDIA's dominant position in AI chips could face increasing pressure. The company's high margins and premium pricing strategy may become harder to maintain as more efficient, lower-cost alternatives emerge. The concentration of value in AI-related growth also creates vulnerability if AI spending patterns shift or if geopolitical tensions impact the global tech supply chain. 
\newline
\newline
While NVIDIA remains a leader in its space, these factors combine to create meaningful downside risks that investors should carefully consider. Despite recent market volatility and concerns about competition from Chinese AI firms like DeepSeek, NVDA maintains strong fundamentals and growth prospects. The company's involvement in Project Stargate, continued innovation in AI hardware, and expansion into new markets like healthcare suggest sustained growth potential. Their dominant market position, strong financial performance, and analyst confidence indicate NVDA remains well-positioned to capitalize on the expanding AI market. 
\newline
\newline
Recent dips in stock price may present buying opportunities for long-term investors, particularly given the company's track record of successful innovation and market leadership in AI technology.
\end{tcolorbox}


\begin{tcolorbox}[
  enhanced,
  colback=gray!5!white,
  colframe=gray!75!black,
  arc=4mm,
  title=Trader Agent \textit{(News only)},
  fonttitle=\bfseries\large,
  coltitle=white,
  colbacktitle=gray!75!black,
  attach boxed title to top right={{xshift=-10mm, yshift=-2mm}}
]
Date: 2025-01-30\\\newline
\newline
After analyzing the news and analyst thoughts, I recommend a SELL decision for NVIDIA stock for the following reasons:
\begin{enumerate}
\item \textbf{Valuation Concerns}: Multiple sources indicate AI chip stocks, including NVIDIA, are trading at ``patently absurd" valuations with triple-digit P/E ratios that appear disconnected from fundamentals, suggesting an overvalued position.
\item \textbf{Emerging Competitive Threats}: The development of China's DeepSeek AI tool represents a tangible competitive threat that has already caused an 8\% pre-market drop in NVIDIA's stock, indicating the market's sensitivity to competitive pressures in the AI chip space.
\item \textbf{AI Spending Bubble Risk}: There are substantial concerns about an AI spending bubble, with analysts warning of an ``inevitable market correction" for AI tech stocks. This suggests NVIDIA's current price may be inflated by unsustainable investment patterns.
\item \textbf{Market Speculation Over Fundamentals}: Multiple articles indicate that NVIDIA and other AI stocks are trading based on speculation (particularly around ``Project Stargate") rather than concrete financial performance, creating vulnerability to sentiment shifts.
\item \textbf{Short-Term Negative Catalysts}: The ``DeepSeek problem" specifically identified for NVIDIA represents an immediate concern that could continue to pressure the stock in the near term.
\end{enumerate}
While NVIDIA maintains a strong position in the AI market with potential long-term government support through initiatives like Project Stargate, the current valuation appears to have priced in overly optimistic scenarios while underweighting significant risks. The combination of excessive valuation, emerging competitive threats, and bubble concerns creates an unfavorable risk-reward profile for the immediate term, making a short position (SELL) the appropriate trading decision for daily position evaluation.\\

Decision: \textbf{SELL}
\end{tcolorbox}

\begin{tcolorbox}[
  enhanced,
  colback=gray!5!white,
  colframe=gray!75!black,
  arc=4mm,
  title=Trader Agent (\textit{News + Risk Model}),
  fonttitle=\bfseries\large,
  coltitle=white,
  colbacktitle=gray!75!black,
  attach boxed title to top right={{xshift=-10mm, yshift=-2mm}}
]
Date: 2025-01-30\\
Model: $\diff S = a(b-S) dt + c\sqrt{S} \diff W$\\
\newline

After analyzing the risk metrics and analyst thoughts, I'm recommending a HOLD position on NVIDIA for the following reasons:

\begin{enumerate}
\item \textbf{Mixed Risk Profile}:
    \begin{itemize}
    \item The risk metrics show some concerning elements: The EVT$_\mathrm{VaR}$ (20.79) and EVT$_\mathrm{CVaR}$ (25.50) are significantly higher than the standard VaR (9.90) and CVaR (16.08), indicating that extreme negative movements could be worse than standard risk measures suggest.
    \item The Maximum Drawdown (MDD) of 52.70\% is substantial and indicates historical vulnerability to major corrections.
    \item However, the negative EVT shape parameters for both losses (-0.15) and residuals (-0.13) suggest bounded tail distributions, which is somewhat reassuring.
    \end{itemize} 
\item \textbf{Neutral Technical Indicators}:
    \begin{itemize}
    \item The RSI at 52.21 is in neutral territory, neither overbought nor oversold.
    \item The positive Drift Score suggests some upward momentum in the price action.
    \item Slightly negative residuals (-0.48) indicate recent underperformance relative to the model's expectations.
    \end{itemize}
\item \textbf{Balanced Fundamental Picture}:
    \begin{itemize}
     \item Strong positives include NVIDIA's dominant AI market position, inclusion in the ``Magnificent 7," and potential benefits from the incoming Trump administration's ``Project Stargate."
    \item Significant concerns include valuation worries, competition from China's DeepSeek AI, and signs of speculation rather than fundamental-driven price action.
   \end{itemize}
\item \textbf{Risk vs. Reward Assessment}:
    \begin{itemize}
   \item While the positive drift suggests upside potential, the elevated tail risk metrics ($\mathrm{EVT}_\mathrm{VaR}$, $\mathrm{EVT}_\mathrm{CVaR}$) indicate substantial downside risk, particularly given the ``AI bubble" concerns mentioned in the analyst thoughts.
   \item The neutral RSI and current market uncertainty suggest waiting for more clarity before taking a directional position.
   \end{itemize}
\end{enumerate}

Given these factors, neither buying nor selling appears clearly advantageous at this moment. The positive drift and NVIDIA's strong market position argue against selling, while the elevated risk metrics and valuation concerns argue against buying. A HOLD recommendation allows for monitoring developments around competitive threats, valuation trends, and ``Project Stargate" specifics before committing to a directional position.\\

Decision: \textbf{HOLD}

\end{tcolorbox}

\section{\texttt{Simudyne Horizon} Synthetic News Context} \label{app:horizon_context}
\subsection*{Summaries of Historic News Events}
Example of historical news headlines:
\begin{tcolorbox}[
  colback=white,
  colframe=black,
  boxrule=1pt,
  arc=0mm,
  boxsep=5pt,
  left=10pt,
  right=10pt,
  top=10pt,
  bottom=10pt,
  enhanced,
  borderline={0.5pt}{0pt}{black}
]

\begin{itemize}
\item (\textit{2022-12-20}) The S\&P 500 edged up marginally by 0.10\% to 3,821.62 on Tuesday, managing to stay in positive territory despite broader global market weakness. Technology stocks showed some resilience, while investors appeared to be cautiously positioning ahead of Thursday's PCE inflation data and Friday's shortened pre-holiday trading session. The slight gain comes as the index continues to navigate concerns about the Federal Reserve's hawkish stance on interest rates heading into 2023. 
\item (\textit{2023-02-02}) The S\&P 500 surged 1.44\% to 4,179.76, extending its rally following the Federal Reserve's widely anticipated 25 basis point rate hike on Wednesday. Technology stocks led the gains as investors interpreted Fed Chair Powell's comments about "disinflation" as a signal that the central bank might be nearing the end of its tightening cycle, despite his caution that more increases would still be needed. 
\item (\textit{2023-03-09}) The S\&P 500 tumbled 1.90\% to 3918.32 on Thursday, marking its worst session in over two weeks as investors grew increasingly concerned about Federal Reserve Chair Jerome Powell's recent hawkish testimony to Congress. Technology and financial stocks led the decline after SVB Financial Group's announcement of a capital raise sparked broader concerns about bank balance sheets and liquidity in a rising rate environment. The labor market data released today showing higher-than-expected jobless claims further fueled investor anxiety about the Fed's aggressive tightening path. 
\item (\textit{2023-04-17}) The S\&P 500 gained 0.32\% on Monday, closing at 4,151.32 as investors showed renewed confidence ahead of key earnings reports from major tech companies this week. Financial and healthcare sectors led the advance, with market participants seemingly shrugging off recent banking sector concerns and focusing on the potential for a less aggressive Fed rate hike path following last week's cooler-than-expected inflation data. 
\item (\textit{2023-05-30}) The S\&P 500 edged slightly higher on Tuesday, gaining a modest 0.05\% to close at 4,205.52 as investors balanced ongoing debt ceiling negotiations with positive consumer confidence data. Technology stocks provided support to the index, with AI-related companies continuing their upward momentum following NVIDIA's recent rally. Market participants remained cautious ahead of upcoming inflation data and the Fed's next policy decision in June. 
\end{itemize}
\end{tcolorbox}

\vspace{1em}
\subsection*{Summaries of Synthetic News Events}
\par Example of synthetic news headlines generated by \texttt{Simudyne Horizon}, which identifies news stories that are consistent with synthetic price paths. Further details are given in \autoref{sec:context}, and can be found at ~\href{https://www.simudyne.com/solutions/simudyne-horizon/}{\texttt{the Simudyne Horizon} product page.}.
\begin{tcolorbox}[
colback=white,
colframe=black,
boxrule=1pt,
arc=0mm,
boxsep=5pt,
left=10pt,
right=10pt,
top=10pt,
bottom=10pt,
enhanced,
borderline={0.5pt}{0.5pt}{black,dotted}
]
\begin{itemize}
\item (\textit{2022-12-20}) The S\&P 500 inched up 0.10\% to 3,821.62 on Tuesday, showing minimal movement as investors remained cautious following the Federal Reserve's hawkish stance last week and amid ongoing concerns about the FTX cryptocurrency collapse's broader financial impact. Energy stocks provided modest support following reports of dwindling U.S. strategic petroleum reserves, while technology shares continued to face pressure as companies reassess spending plans in light of Amazon's recent massive layoff announcement. Market participants appeared hesitant to make significant moves heading into the holiday-shortened trading week, with many awaiting tomorrow's consumer confidence data for further insights into recession risks.
\item (\textit{2023-02-02})  The S\&P 500 plunged 1.62\% to 3,796.29 on Thursday as markets reeled from the widespread effects of a major cyber attack targeting global financial payment systems, now in its third day. Financial stocks led the decline as major banks and payment processors continued to struggle with service disruptions, while tech companies involved in cybersecurity saw mixed performance amid questions about their ability to contain the breach. Investor sentiment deteriorated further following testimony from Treasury officials suggesting the attack could have longer-term implications for financial stability than initially anticipated.
\item (\textit{2023-03-09})  The S\&P 500 managed a modest gain of 0.51\% to close at 3,800.86 on Thursday, showing resilience despite ongoing concerns about the banking sector following Credit Suisse's collapse earlier this week. Technology stocks led the rebound as investors cautiously returned to growth sectors after the Supreme Court's AI licensing decision provided some regulatory clarity, though financial stocks remained under pressure as market participants continued to assess potential systemic risks from the banking crisis. The index's performance suggests investors are beginning to look past immediate banking concerns while remaining cautious about the broader economic implications of recent financial sector turmoil.
\item (\textit{2023-04-17})  The S\&P 500 tumbled 0.96\% to 3,934.62 on Monday as investors reacted to the Federal Reserve's unexpected emergency meeting announcement scheduled for tomorrow, fueling speculation about potential interest rate actions outside the regular schedule. Banking stocks led the decline following reports that additional regional banks are facing liquidity pressures despite the recent Credit Suisse rescue, while technology shares were further pressured by ongoing uncertainty surrounding the Supreme Court's recent AI licensing requirements that continue to create implementation challenges for major tech firms.
\item (\textit{2023-05-30})  The S\&P 500 edged up 0.25\% to 4,226.27 on Tuesday, extending its cautious recovery as investors digested the recently announced bipartisan corporate tax reform proposal that has boosted market sentiment. Energy stocks led gains following rumors of verification data supporting last week's major North American oil discovery announcement, while technology shares showed mixed performance as the market continues to process the implications of the Supreme Court's AI licensing requirements. Healthcare stocks remained under pressure as analysts revised growth forecasts downward in the wake of last year's blocked pharmaceutical merger.
\end{itemize}
\end{tcolorbox}

\end{document}